\pdfoutput=1

\documentclass[11pt,twoside,a4paper,cmspaper,final,collab]{cms-tdr}
\def\svnVersion{490635P}\def\cmsCernNoTag{CERN-EP-2018-218}\def\cmsCernDate{\today}\def\cmsMessage{Published in Physical Review D as \href{http://dx.doi.org/10.1103/PhysRevD.99.032014}{\doi{10.1103/PhysRevD.99.032014.}}}

\begin{document}\cmsNoteHeader{EXO-17-003}

\hyphenation{had-ron-i-za-tion}
\hyphenation{cal-or-i-me-ter}
\hyphenation{de-vices}
\RCS$HeadURL: svn+ssh://svn.cern.ch/reps/tdr2/papers/EXO-17-003/trunk/EXO-17-003.tex $
\RCS$Id: EXO-17-003.tex 489119 2019-02-14 21:05:03Z dmorse $
\newlength\cmsFigWidth
\ifthenelse{\boolean{cms@external}}{\setlength\cmsFigWidth{0.49\textwidth}}{\setlength\cmsFigWidth{0.75\textwidth}}
\ifthenelse{\boolean{cms@external}}{\providecommand{\cmsLeft}{upper\xspace}}{\providecommand{\cmsLeft}{left\xspace}}
\ifthenelse{\boolean{cms@external}}{\providecommand{\cmsRight}{lower\xspace}}{\providecommand{\cmsRight}{right\xspace}}
\ifthenelse{\boolean{cms@external}}{\providecommand{\cmsLL}{middle\xspace}}{\providecommand{\cmsLL}{lower left\xspace}}
\ifthenelse{\boolean{cms@external}}{\providecommand{\cmsLR}{lower\xspace}}{\providecommand{\cmsLR}{lower right\xspace}}
\ifthenelse{\boolean{cms@external}}{\providecommand{\NA}{\ensuremath{\cdots}\xspace}}{\providecommand{\NA}{\ensuremath{\text{---}}\xspace}}
\ifthenelse{\boolean{cms@external}}{\providecommand{\CL}{C.L.\xspace}}{\providecommand{\CL}{CL\xspace}}
\providecommand{\cmsTable}[1]{\resizebox{\textwidth}{!}{#1}}
\newcommand{\mumujj}{\ensuremath{\PGm\PGm\mathrm{jj}}}
\newcommand{\munujj}{\ensuremath{\PGm\Pgn\mathrm{jj}}}
\newcommand{\st}{\ensuremath{S_{\mathrm{T}}^{\PGm\PGm\mathrm{jj}}}}
\newcommand{\Muu}{\ensuremath{m_{\PGm\PGm}}}
\newcommand{\Mujmin}{\ensuremath{m_{\PGm\mathrm{j}}^{\mathrm{min}}}}
\newcommand{\Muj}{\ensuremath{m_{\PGm\mathrm{j}}}}
\newcommand{\stuv}{\ensuremath{S_{\mathrm{T}}^{\PGm\Pgn\mathrm{jj}}}}
\newcommand{\mt}{\ensuremath{m_{\mathrm{T}}^{\PGm\Pgn}}}
\newcommand{\mlq}{\ensuremath{m_{\mathrm{LQ}}\xspace}}
\newlength\cmsTabSkip\setlength{\cmsTabSkip}{1ex}

\cmsNoteHeader{EXO-17-003}
\title{Search for pair production of second-generation leptoquarks at \texorpdfstring{$\sqrt{s}=13\TeV$}{sqrt(s) = 13 TeV}}

\date{\today}

\abstract{A search for pair production of second-generation leptoquarks is performed using proton-proton collision data collected at $\sqrt{s}=13\TeV$ in 2016 with the CMS detector at the CERN LHC, corresponding to an integrated luminosity of 35.9\fbinv. Final states with two muons and two jets, or with one muon, two jets, and missing transverse momentum are considered. Second-generation scalar leptoquarks with masses less than 1530\,(1285)\GeV are excluded for $\beta = 1.0$\,(0.5), where $\beta$ is the branching fraction for the decay of a leptoquark to a charged lepton and a quark. The results of the search are also interpreted as limits on the pair production of long-lived top squarks in an $R$-parity violating supersymmetry model that has a final state with two muons and two jets. These limits represent the most stringent limits to date on these models.}

\hypersetup{
pdfauthor={CMS Collaboration},
pdftitle={Search for pair production of second-generation leptoquarks at sqrt(s)=13 TeV},
pdfsubject={CMS},
pdfkeywords={CMS, physics, leptoquark, muon, jet, exotica}}

\maketitle

\section{Introduction}
\label{sec:introduction}

The standard model (SM) of particle physics displays a symmetry between the quark and lepton families. Leptoquarks (LQs) are new bosons that would manifest a fundamental connection between quarks and leptons and are predicted by numerous extensions of the SM, such as grand unified theories~\cite{gut0,gut1,gut2,gut3,gut4,gut5,gut6,gut7}, composite models with lepton and quark substructure~\cite{composite}, technicolor models~\cite{technicolor1,technicolor2,technicolor3}, and superstring-inspired models~\cite{superstring_e6}. LQs are color-triplet scalar or vector bosons carrying both lepton and baryon numbers, and decay either to a charged lepton and a quark, or to a neutrino and a quark. Interpretations of direct searches for LQs are typically based on a general model where LQ-lepton-quark interactions are added to the Lagrangian~\cite{mBRW}. Recently, interest in LQs has increased as they may provide an explanation for the observation of anomalies in the decays of \PB{} mesons by the Belle~\cite{Matyja:2007kt,Bozek:2010xy,Huschle:2015rga}, \textsc{BaBar}~\cite{Lees:2012xj,Lees:2013uzd}, and LHCb~\cite{Aaij:2015yra,Aaij:2013qta,Aaij:2014ora,Aaij:2017vbb} Collaborations.

\begin{figure*}[htbp]
       \centering
       {\includegraphics[width=.3\textwidth]{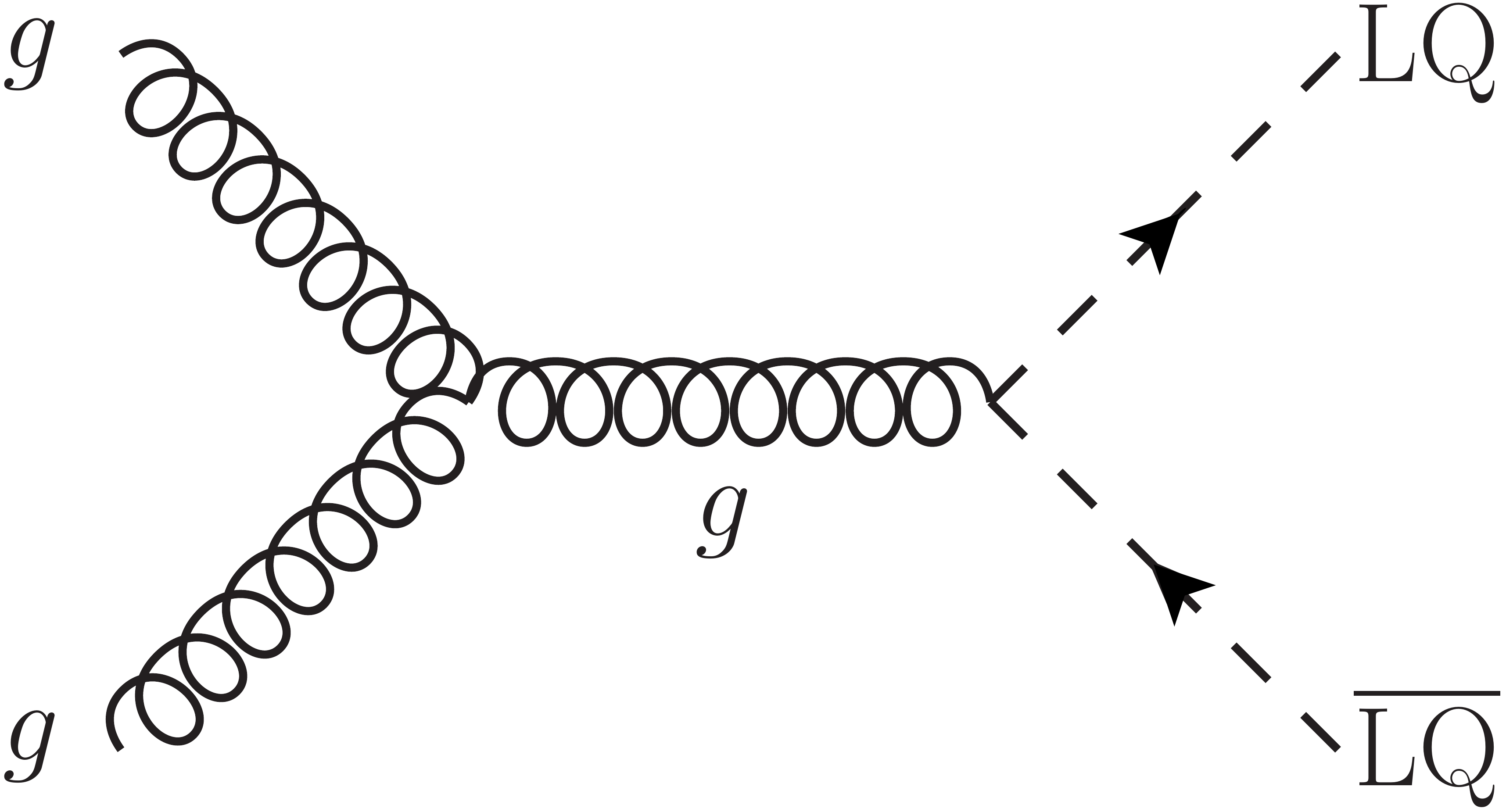}}\hspace*{0.4cm}
       {\includegraphics[width=.3\textwidth]{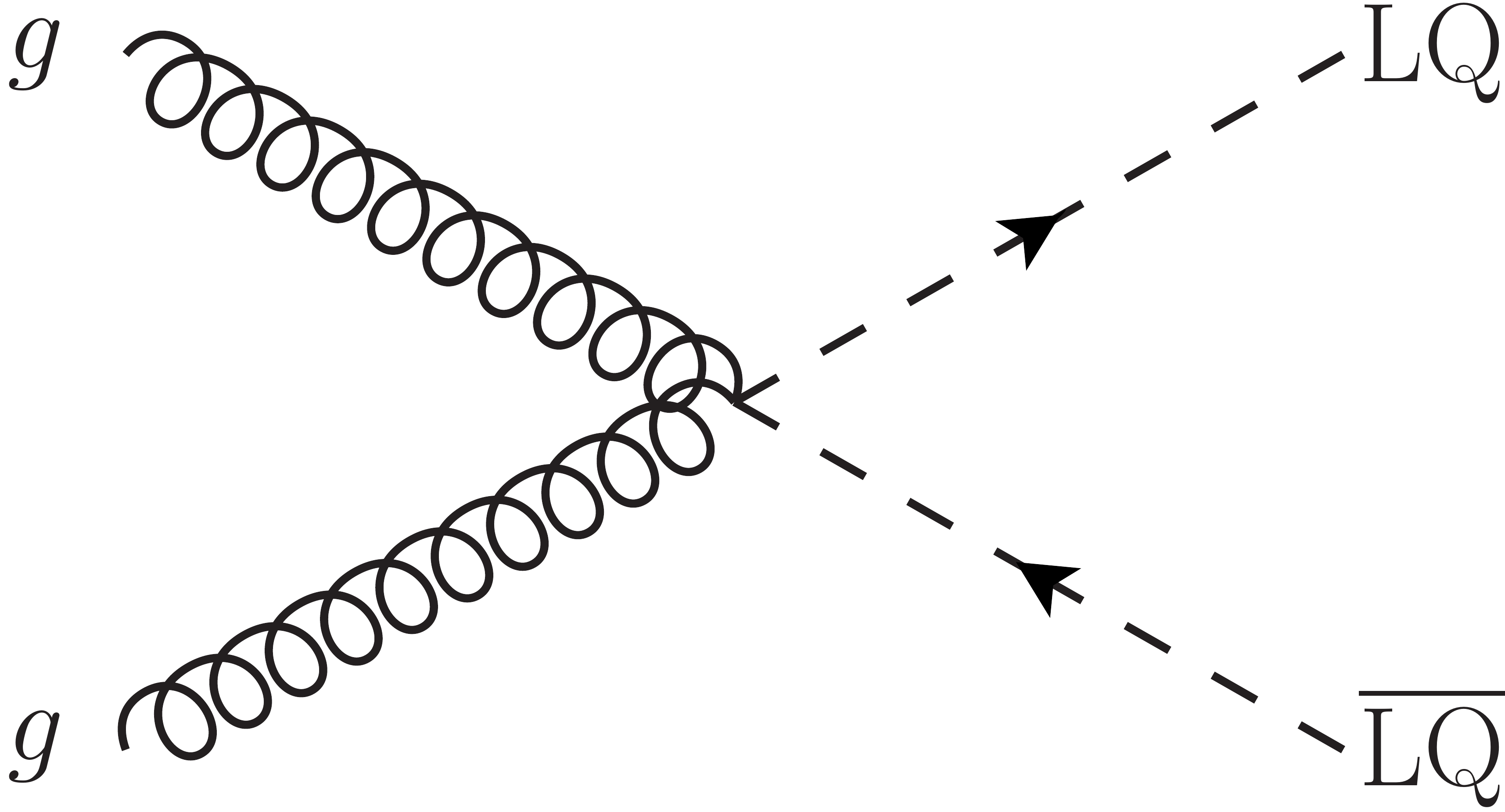}}\hspace*{0.4cm}
       {\includegraphics[width=.3\textwidth]{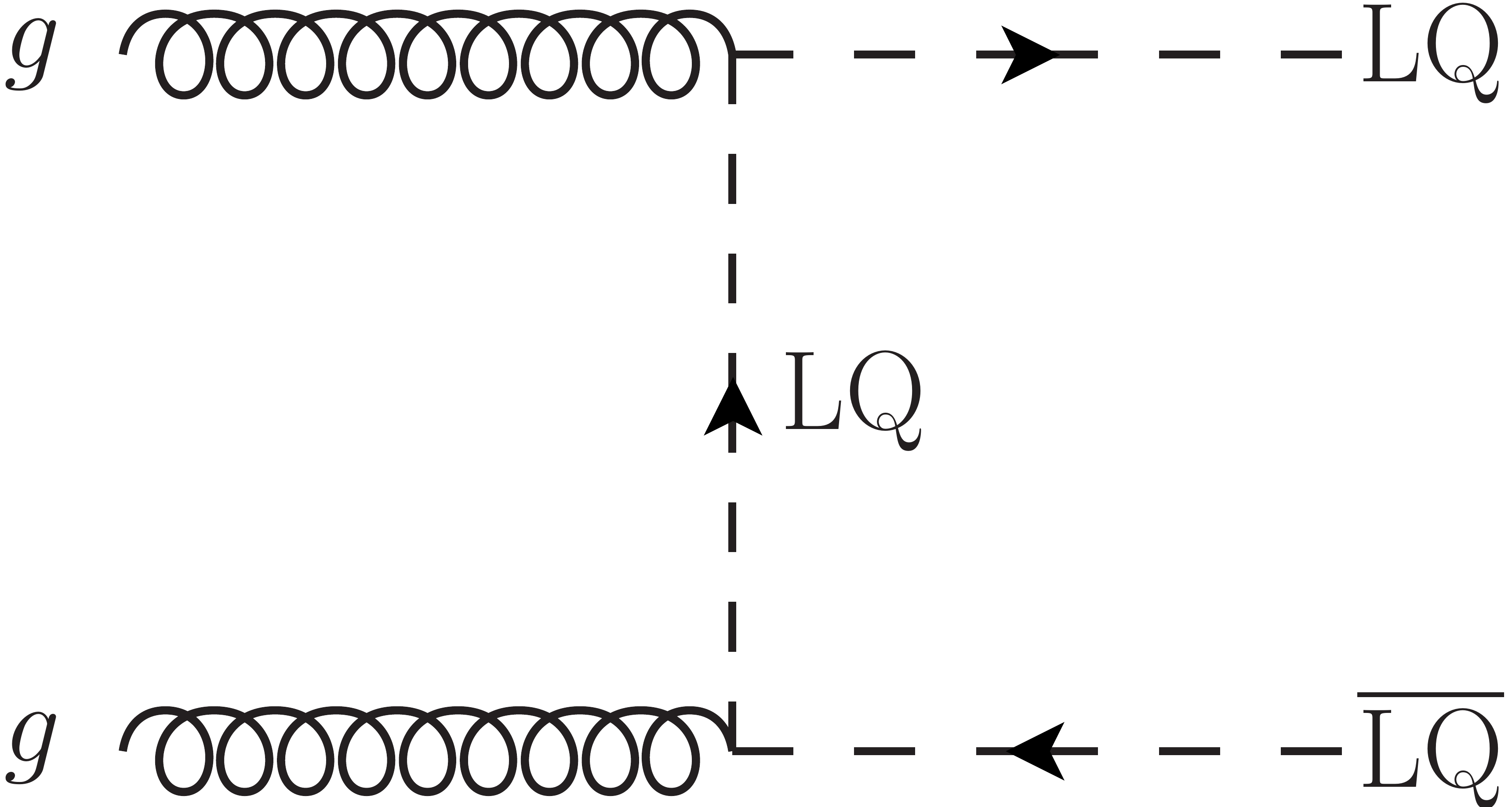}}\\
       \vspace{0.5cm}
       {\includegraphics[width=.3\textwidth]{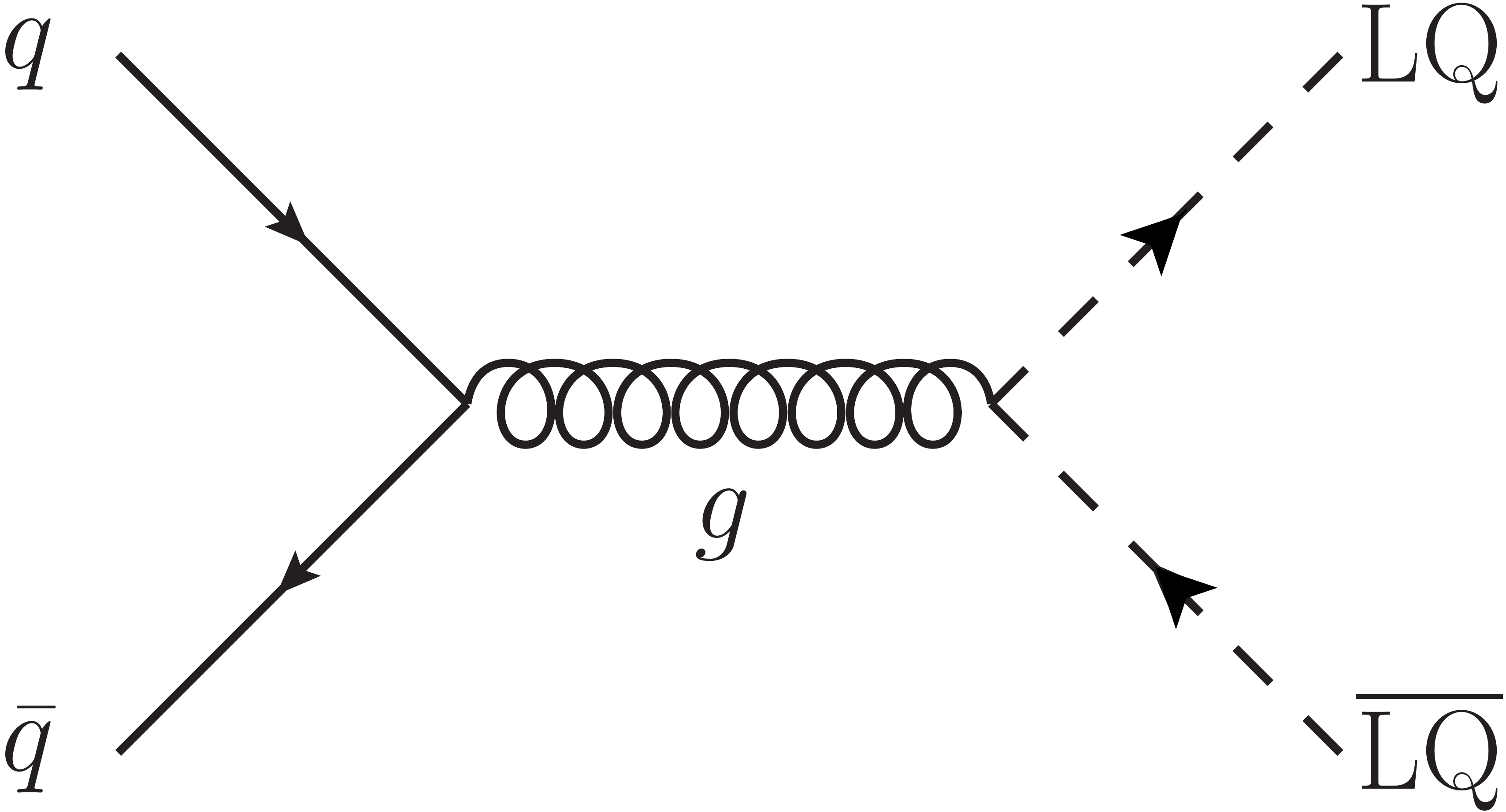}}
       \caption{Dominant leading-order Feynman diagrams for the pair production of LQs at the LHC.}
          \label{fig:LQFeynman}
\end{figure*}

 At hadron colliders, LQs can be produced singly or in pairs. This analysis concentrates on pair production of scalar LQs. The dominant leading-order (LO) processes for pair production of LQs at the LHC involve gluon-gluon fusion and quark-antiquark annihilation, shown in Fig.~\ref{fig:LQFeynman}.

The interactions of scalar LQs with SM particles are completely determined by three parameters~\cite{mBRW}: the LQ mass {\mlq}, the Yukawa coupling at the LQ-lepton-quark vertex $\lambda_{\mathrm{LQ}}$, and the branching fraction $\beta$ of the LQ decay to a charged lepton and a quark. The decay of an LQ to a neutrino and a quark is complementary to the decay to a charged lepton and a quark and has a branching fraction of $1-\beta$. Vector LQs are further dependent on two couplings which relate to the anomalous magnetic and electric quadrupole moments of the vector LQ~\cite{Blumlein:1996qp}.

As can be seen in Fig.~\ref{fig:LQFeynman}, the dominant pair production processes have no LQ-lepton-quark vertices, and thus the production cross sections do not depend on $\lambda_{\mathrm{LQ}}$. The mean lifetime of the LQ is dependent on $\lambda_{\mathrm{LQ}}$. For $\lambda_{\mathrm{LQ}}\gtrsim10^{-6.5}$~\cite{mBRW}, {\TeV}-scale LQs will have decay lengths that are less than the resolution of the impact parameter measurement of the CMS detector~\cite{Diaz:2017lit}. As is customary, the value of $\lambda_{\mathrm{LQ}}$ has been set such that $\lambda_{\mathrm{LQ}}^{2}/(4\pi) = \alpha_{\mathrm{em}}$, where  $\alpha_{\mathrm{em}}$ is the electromagnetic coupling. Therefore the LQs considered in this analysis always decay very close to the point of production and are referred to as prompt. As a consequence, the limits set on the pair production cross sections can be considered independent of $\lambda_{\mathrm{LQ}}$ for $\lambda_{\mathrm{LQ}}\gtrsim10^{-6.5}$.

Pair production of LQs is characterized by final states with two leptons and two jets with large transverse momentum \pt. This analysis assumes no flavor mixing between generations, to be consistent with experimental constraints on lepton flavor violation and flavor-changing neutral currents~\cite{lq_constraints,FCNC}. In this scenario, second-generation LQs will always decay to either a muon and a charm quark, or to a neutrino and a strange quark. Values of 1.0 and 0.5 are considered for $\beta$, corresponding to maximal production of the two final states {\mumujj} and {\munujj}. Previous limits on second-generation scalar LQ pair production have been published by the CMS and ATLAS Collaborations~\cite{Khachatryan:2015vaa,Aaboud:2016qeg}. The CMS result excludes LQs with $\mlq < 1080$\,(760)\GeV for $\beta = 1.0$\,$(0.5)$, in proton-proton ($\Pp\Pp$) collisions at 8\TeV, and ATLAS excludes LQs with $\mlq < 1160\GeV$ for $\beta=1.0$, at 13\TeV. The most stringent limits on vector LQs have been reported by CMS~\cite{Khachatryan:2015vaa}.

Other models of physics beyond the SM, such as $R$-parity violating (RPV) supersymmetry (SUSY)~\cite{rpv}, can lead to the same final states as LQ production. Supersymmetry postulates a symmetry between fermions and bosons, which gives rise to superpartner particles for all known SM particles. In some SUSY scenarios, one of the two top quark superpartners (top squark, \PSQt) is the lightest SUSY particle and when $R$-parity is violated can decay to a bottom (\PQb) quark and a charged lepton. For {\PSQt} pair production and direct {\PSQt} decays to charged lepton + \PQb{} quark, limits can be extracted directly from the LQ results. If the couplings of the RPV operators are sufficiently small, however, the superpartners will have long lifetimes, and will travel through part or all of the detector before decaying. In this scenario, referred to in this paper as displaced SUSY~\cite{dispSUSY}, the {\PSQt} has a finite but non-zero lifetime, and decays to a charged lepton of any flavor and a bottom quark within a distance, $c\tau$, between 0.1 and 100\unit{cm}, where {$\tau$} is the {\PSQt} mean lifetime. We assume the {\PSQt} decays with equal probability to electrons, muons, and tau leptons. This analysis is sensitive to the low-lifetime, high-mass region of phase space where dedicated searches for displaced SUSY lose sensitivity~\cite{Khachatryan:2014mea}.

\section{The CMS detector}
\label{sec:CMS}
The central feature of the CMS apparatus is a superconducting solenoid of 6\unit{m} internal diameter, providing a magnetic field of 3.8\unit{T}. Within the solenoid volume are a silicon pixel and strip tracker, a lead tungstate crystal electromagnetic calorimeter (ECAL), and a brass and scintillator hadron calorimeter, each composed of a barrel and two endcap sections. Forward calorimeters extend the pseudorapidity coverage provided by the barrel and endcap detectors. Muons are measured in gas-ionization detectors embedded in the steel flux-return yoke outside the solenoid. A more detailed description of the CMS detector, together with a definition of the coordinate system used and the relevant kinematic variables, can be found in Ref.~\cite{cms-jinst}.

Events of interest are selected using a two-tiered trigger system~\cite{Khachatryan:2016bia}. The first level (L1), composed of custom hardware processors, uses information from the calorimeters and muon detectors to select events at a rate of around 100\unit{kHz} within a time interval of less than 4\mus. The second level, known as the high-level trigger (HLT), consists of a farm of processors running a version of the full event reconstruction software optimized for fast processing, and reduces the event rate to around 1\unit{kHz} before data storage.

\section{Data and simulated samples}
\label{sec:samples}

The data set used in this paper was collected by CMS during the 2016 {\Pp\Pp} LHC run at $\sqrt{s} = 13\TeV$ and corresponds to an integrated luminosity of $35.9 \pm 0.9$\fbinv~\cite{lumi}. Events are selected using triggers that require at least one muon with $\pt > 50\GeV$, with no isolation requirements. These triggers supply the data for the {\mumujj} and {\munujj} channels, as well as for the {\Pe\PGm} sample used in the {\ttbar}+jets background estimate for the {\mumujj} channel.

Signal samples are produced in 50\GeV steps for scalar {\mlq} between 200 and 2000\GeV using an effective theory based on Ref.~\cite{mBRW} at LO with \textsc{pythia} 8.212~\cite{pythia8p2}. These samples are used to study the acceptance of the signal. The production cross sections, calculated using next-to-leading order (NLO) QCD corrections~\cite{kramer} with the \textsc{cteq6l1}~\cite{cteq6l1} LO and \textsc{cteq6.6}~\cite{CTEQ6p6} NLO PDF sets, are used for comparison with data in the limit setting procedure. The search limits are independent of $\lambda_{\mathrm{LQ}}$ for sufficiently large values of $\lambda_{\mathrm{LQ}}$, as discussed in Section~\ref{sec:introduction}. Displaced SUSY samples are produced with \textsc{pythia} 8.212 using the Snowmass ``Points and Slopes point 1a'' parameter set~\cite{Allanach:2002nj} for {\PSQt} masses from 200 to 1200\GeV, in 100\GeV steps, and for $c\tau = 0.1$, 1, 10, and 100\unit{cm}. The lighter, left-handed top squark is the lightest supersymmetric particle in this model, while the heavier right-handed top squark has a mass beyond the relevant kinematic regime. Production cross sections for {\PSQt} are calculated at NLO + next-to-leading logarithmic (NLL) precision with \textsc{prospino} version 2~\cite{prospino} and NLL-fast programs version 3.0~\cite{NLLfast1,NLLfast2}, using the \textsc{cteq6l1} PDF set.

Standard model backgrounds considered include $\cPZ/\gamma^*$+jets, {\ttbar}+jets, $\PW$+jets, single top quark production, and diboson ($\PW\PW$/$\PW\cPZ$/$\cPZ\cPZ$)+jets. The $\cPZ/\gamma^*$+jets, $\PW$+jets, and diboson samples are generated at NLO using {\MGvATNLO} version 2.3.3~\cite{mg5_amcnlo,amcnlo}. Single top quark and {\ttbar}+jets samples are generated at NLO using \textsc{powheg} v2~\cite{Frixione:2007vw,Alioli:2010xd,powheg,Alioli:2009je} and {\MGvATNLO} \cite{Frixione:2005vw}. All backgrounds use \textsc{pythia} 8.212 for fragmentation and hadronization.

The $\PW$+jets and $\cPZ/\gamma^*$+jets samples are normalized to next-to-next-to-leading order (NNLO) inclusive cross sections calculated with {\FEWZ} versions 3.1 and 3.1.b2, respectively~\cite{FEWZ}. Single top quark and diboson samples are normalized to NLO inclusive cross sections calculated with {\MCFM} version 6.6~\cite{ mcfm_t, mcfm_Wt, mcfm_tch,mcfm_diboson}. The {\ttbar}+jets sample is normalized to calculations at the NNLO level in QCD including resummation of next-to-next-to-leading logarithmic (NNLL) soft gluon terms produced with Top++2.0~\cite{tt1,tt2,tt3,tt4,tt5,tt6,tt7}.

Signal and background events are generated using the \textsc{nnpdf3.0} parton distribution function (PDF) sets~\cite{nnpdf}, with the full CMS detector geometry and response simulated using {\GEANTfour}~\cite{Geant1,Geant2}. All samples use the \textsc{cuetp8m1} underlying event tune~\cite{Khachatryan:2015pea}, with additional {\Pp\Pp} interactions (the pileup distribution) overlaid  and corrected to match the distribution measured in data.

The simulated samples are corrected so that the detector response and resolution for both leptons and jets and the triggering efficiency match those measured in data.

\section{Event reconstruction and selection}
\label{sec:eventselection}

The CMS particle-flow event algorithm~\cite{particle_flow} aims to reconstruct and identify each individual particle in an event, with an optimized combination of information from the various elements of the detector. The reconstructed vertex with the largest value of summed physics-object $\pt^2$ is taken to be the primary {\Pp\Pp} interaction vertex. The physics objects in this context are jets clustered using the jet finding algorithm with all tracks assigned to the vertex as inputs, and the associated missing transverse momentum \ptvecmiss, taken as the negative vector sum of the {\pt} of those jets. The magnitude of the {\ptvecmiss} is referred to as \ptmiss.

Hadronic jets are reconstructed using the anti-\kt algorithm~\cite{antiktjets,Cacciari:2011ma} with a size parameter of 0.4. Jet momentum is determined as the vectorial sum of all particle momenta in the jet, and is found from simulation to be within 5 to 10\% of the true momentum over the whole \pt spectrum and detector acceptance. Additional {\Pp\Pp} interactions within the same or nearby bunch crossings can contribute additional tracks and calorimetric energy depositions, increasing the apparent jet momentum. To mitigate this effect, tracks identified to be originating from pileup vertices are discarded, and an offset correction is applied to correct for remaining contributions. Jet energy corrections are derived from simulation to bring the measured response of jets to that of particle level jets on average. In situ measurements of the momentum balance in dijet, photon+jet, \cPZ+jet, and multijet events are used to determine any residual differences between jet energy scale in data and in simulation and appropriate corrections are made~\cite{Khachatryan:2016kdb}. These jet energy corrections are propagated to the \ptmiss. Additional selection criteria are applied to each jet to remove jets potentially dominated by instrumental effects or reconstruction failures. Jets are required to have pseudorapidity $\abs{\eta} < 2.4$, $\pt > 50\GeV$, and to be separated from all selected muons by {$\Delta R > 0.5$}, where {$\Delta R = \sqrt{(\Delta\eta)^2+(\Delta\phi)^2}$} and $\phi$ is the azimuthal angle in radians. At least two jets are required for both the {\mumujj} and {\munujj} channels, with no jet flavor requirement. Jets originating from \PQb{} quarks are used to estimate backgrounds in data control regions, and are identified using the combined secondary vertex algorithm~\cite{btag}. Jets are considered as \PQb-tagged if they pass the `loose' working point, with an 80\% \PQb{} jet identification efficiency and a 10\% rate of erroneous \PQb{} jet identification. Simulated samples are corrected on a jet-by-jet basis using correction factors to agree with \PQb-tagged distributions measured in data.

Muons are measured in the pseudorapidity range $\abs{\eta} < 2.4$ in concentric stations with detection planes made using three technologies: drift tubes, cathode strip chambers, and resistive plate chambers.  Hits in the muon tracking system are combined into hit segments. Muons are reconstructed as tracks by combining these hit segments with hits in the silicon tracker, with a reconstruction optimized for high {\pt} muons. Matching muons to tracks measured in the silicon tracker results in a relative {\pt} resolution for muons with $\pt < 100\GeV$ of 1\% in the barrel and 3\% in the endcaps. The {\pt} resolution in the barrel and endcaps is better than 10\% for muons with {\pt} up to 1\TeV~\cite{MuId}. Muons are required to have $\pt > 53\GeV$ and $\abs{\eta} < 2.4$ to be fully efficient with respect to the trigger, and are required to satisfy a set of identification criteria optimized for high {\pt}. Segments in at least two muon stations are required to be geometrically matched to a track in the silicon tracker, with at least one hit from a muon chamber in each station included in the muon track fit. In order to suppress muons from hadron decays and to allow for a more precise {\pt} measurement, at least five strip tracker layers with hits associated with the muon are required, and at least one hit in the pixel detector. To reject muons from cosmic rays, the transverse impact parameter of the muon track with respect to the primary vertex is required to be less than 2\unit{mm}, and the longitudinal distance of the track with respect to the primary vertex is required to be less than 5\unit{mm}. An isolation requirement is imposed, as the signal produces isolated muons. The {\pt} sum of all tracks from the primary vertex (excluding the muon track itself) in a cone of {$\Delta R = 0.3$} around the muon track, divided by the muon {\pt}, is required to be less than 0.1. This relative isolation is shown to be independent of pileup~\cite{MuId}. In the {\mumujj} channel at least two muons are required, with no charge requirement. In the {\munujj} channel exactly one muon is required.

Electrons are measured in the pseudorapidity range $\abs{\eta} < 2.5$. The electron momentum is estimated by combining the energy measurement in the ECAL with the momentum measurement in the tracker. The momentum resolution for electrons with $\pt \approx 45\GeV$ from $\cPZ\to\Pe\Pe$ decays ranges from 1.7\% to 4.5\%~\cite{Khachatryan:2015hwa}. In this analysis electrons are used as a control data sample for a {\ttbar}+jets background estimate in the {\mumujj} channel, and electrons with $\pt > 53\GeV$ are vetoed in the {\munujj} channel to avoid overlap with this control region. With this high veto threshold the selection is kept as inclusive as possible for signal. The {\ttbar}+jets background is small in the mass region of interest, which is above 1\TeV.

The LQ candidates are reconstructed using the pairing where the two reconstructed masses are closest. In the {\mumujj} channel the two highest {\pt} muons and two highest {\pt} jets that pass the selection criteria above are considered. Each muon is paired with a jet in the configuration that minimizes the LQ-$\overline{\mathrm{LQ}}$ invariant mass difference. In the {\munujj} channel the two highest {\pt} jets are considered together with the required single muon. The muon and {\ptvecmiss} are each paired with a jet in a similar manner to the {\mumujj} channel, using instead the LQ transverse masses $m_{\mathrm T}^{\mathrm{LQ}}=\sqrt{\smash[b]{2\pt^{\ell}\pt^{\text{jet}}(1-\cos[\Delta\phi(\ell,\text{jet})])}}$ of the muon-jet and \ptvecmiss-jet systems, where in this case {$\ell$} represents the muon or neutrino in the decay. This method correctly matches the decay products of the two LQs in 50 to 70\% of signal events, increasing with \mlq.

\section{Estimation of standard model backgrounds}
\label{sec:back}

\subsection{\texorpdfstring{The {\mumujj} channel}{The mumujj channel}}
\label{sec:back:mumu}
The main backgrounds that can mimic the LQ signal in the {\mumujj} channel are $\cPZ/\gamma^*$+jets and {\ttbar}+jets events.

Backgrounds are estimated and validated using a selection dominated by background events, referred to as the preselection. The preselection applies criteria that are looser than any final selection. This preselection requires at least two muons with $\pt > 53\GeV$ and at least two jets with $\pt > 50\GeV$. The muons are required to be separated from one another by $\Delta R > 0.3$. The invariant mass of the dimuon system (\Muu) is required to be greater than 50\GeV, and the {\st} of the event is required to be greater than 300\GeV, where {\st} is defined as the scalar sum of the {\pt} of the two jets and two muons in the event.

The $\cPZ/\gamma^*$+jets background is estimated with events that satisfy the preselection, in a data control region around the {\cPZ} peak that is not in the search region. The background shape is taken from simulation, which shows good shape agreement with the data in the control region. For normalization, the simulation is compared to data in a window $80 < M_{\Pgm\Pgm} < 100\GeV$ around the {\cPZ} peak, and a measured data normalization scale factor of $0.98\pm 0.01\stat\pm 0.09\syst$ is applied to simulated events passing the final selection criteria. The systematic uncertainty is assigned to account for the dependence of the scale factor on event kinematic properties. All final selections require $M_{\Pgm\Pgm} > 100\GeV$, to reduce the $\cPZ/\gamma^*$ background, and to maintain the separation of the control region from the search region.

The {\ttbar}+jets background is estimated using an independent {\Pe\Pgm} data sample. Events are selected that contain one electron and one muon, and must satisfy all requirements of the {\mumujj} preselection, other than the normal two muon requirement. No charge requirement is placed on the electron and muon. This sample is corrected for differences between the {\Pgm\Pgm} and {\Pe\Pgm} selection, such as those based on identification and isolation, as well as on trigger efficiency. The kinematic distributions of this sample are found to be in good agreement with the {\ttbar}+jets simulation, and use of the {\Pe\Pgm} control sample in data reduces the systematic uncertainties associated with this background.

Background contributions from single top quark, $\PW$+jets, and diboson events are estimated from simulation. Background from QCD multijets is shown to be negligible using data control regions.

\begin{figure}[htb]
       \centering
       {\includegraphics[width=.49\textwidth]{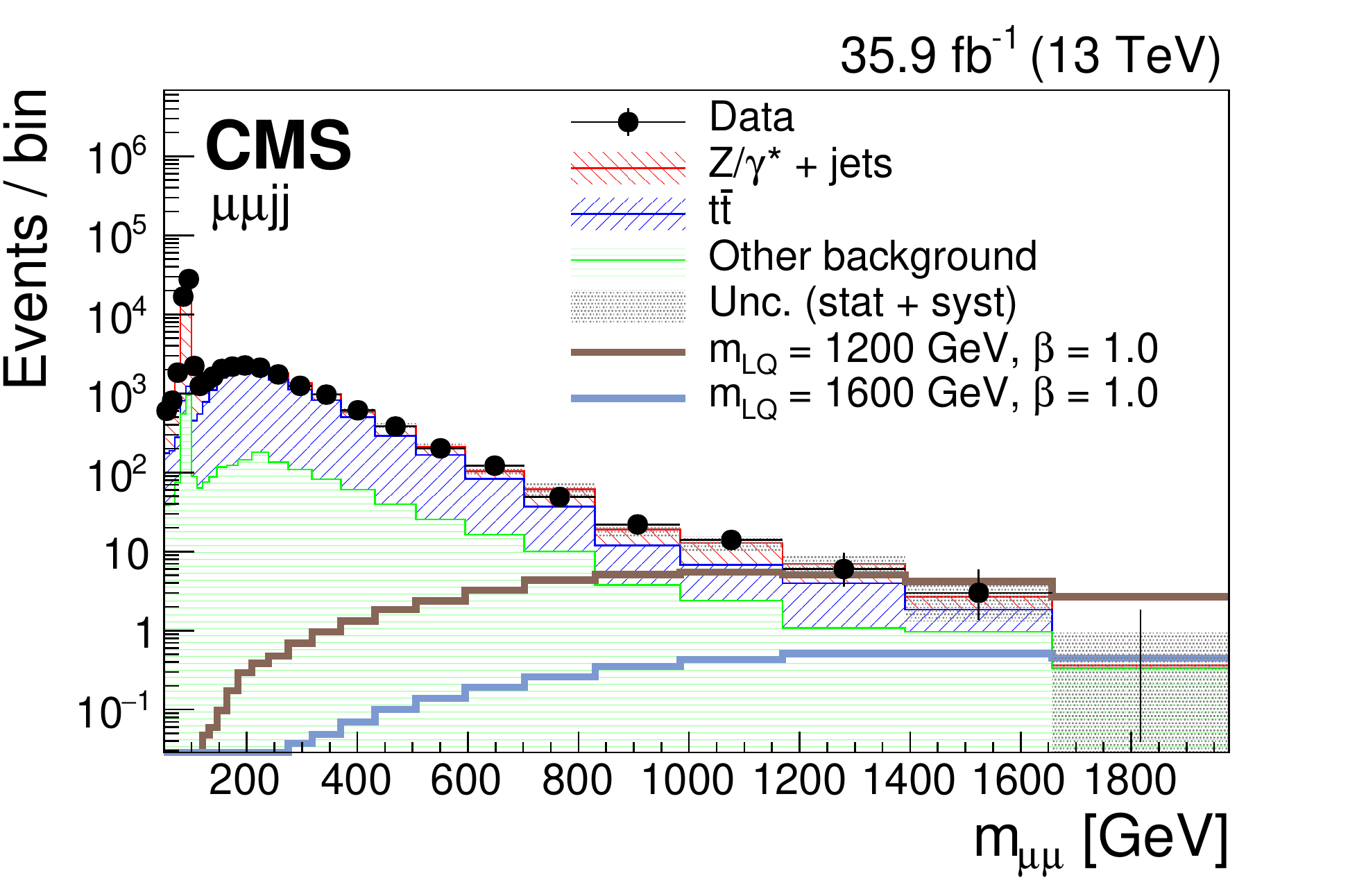}}\\
       {\includegraphics[width=.49\textwidth]{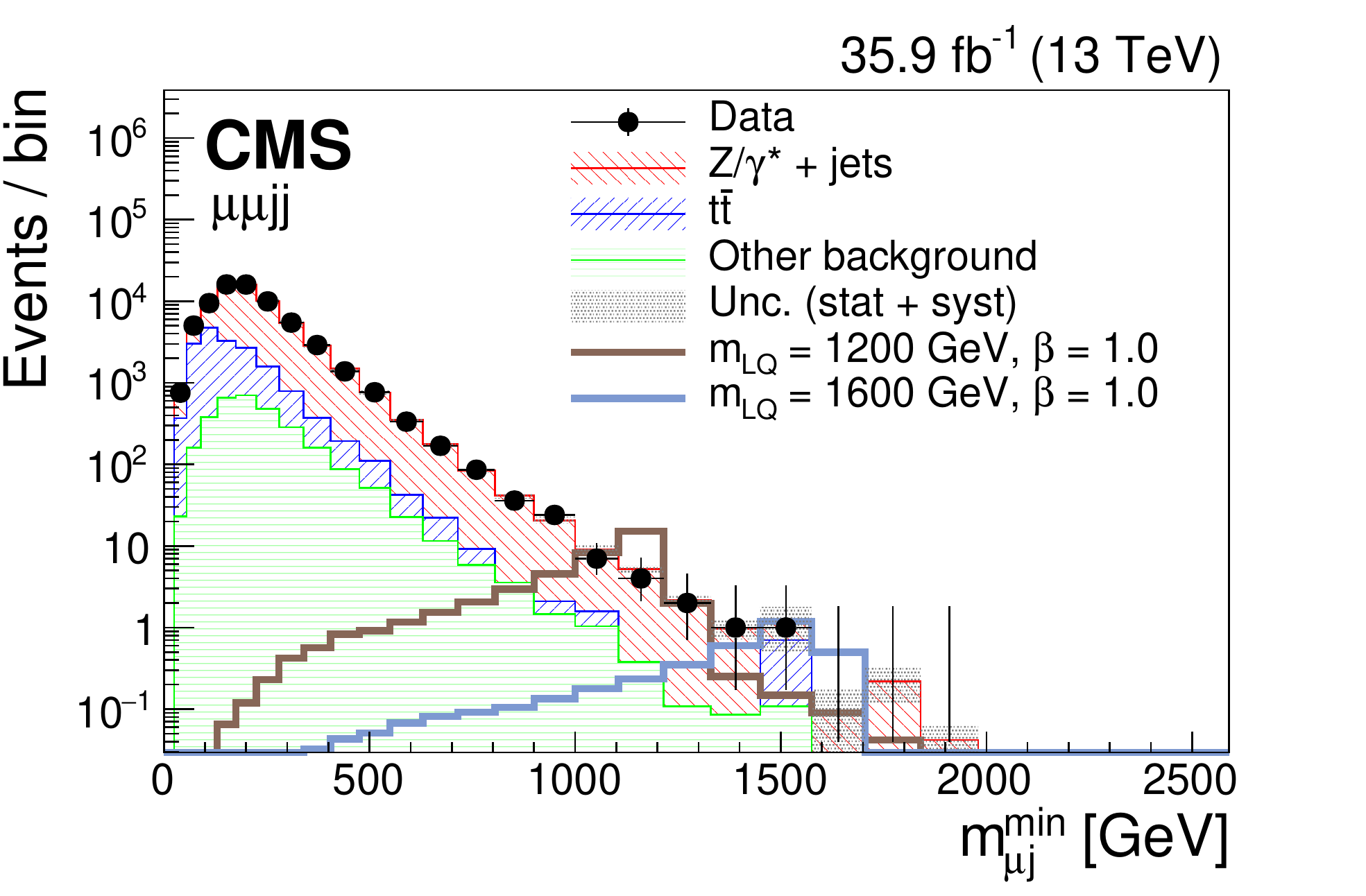}}
       {\includegraphics[width=.49\textwidth]{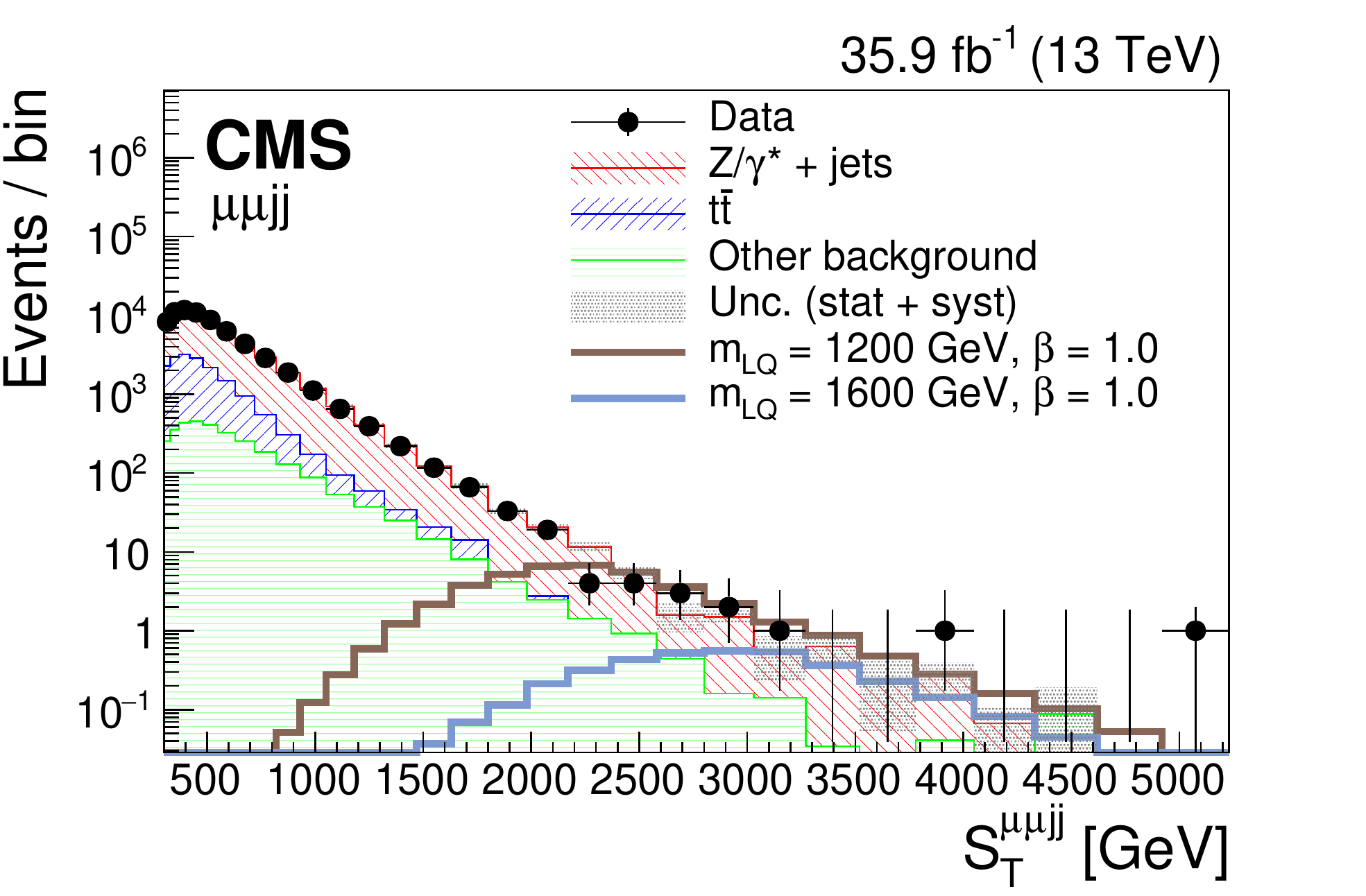}}
       \caption{Comparison of data and background at the preselection level for the {\mumujj} channel, for the variables used for final the selection optimization: {\Muu} (upper), {\Mujmin} (\cmsLL), and {\st} (\cmsLR). `Other background' includes $\PW$+jets, single top quark, and diboson backgrounds. The hashed band represents the combined statistical and systematic uncertainty in the full background estimate.}
          \label{fig:LQMumu}
\end{figure}

Background predictions are validated at the preselection level by comparing them with data. Good agreement is seen in all relevant kinematic distributions. Three kinematic variables are identified that have strong discrimination power between signal and background. In the {\mumujj} channel, these variables are \st, \Muu, and \Mujmin, where {\Mujmin} is defined as the smaller of the two muon-jet invariant masses that represent the LQ and $\overline{\mathrm{LQ}}\xspace$ candidates. A comparison of these main kinematic variables is shown in Fig.~\ref{fig:LQMumu}
at the preselection level.

\subsection{\texorpdfstring{The {\munujj} channel}{The munujj channel}}
\label{sec:back:munu}
As in the {\mumujj} channel, a background-dominated preselection is used to calculate and validate the SM background estimates. This preselection requires exactly one muon with $\pt > 53\GeV$ and at least two jets with $\pt > 50\GeV$. The direction of the muon in the event is required to be separated from {\ptvecmiss} by $\Delta\phi > 0.8$, and the momentum vector of the highest-{\pt} jet to be separated from {\ptvecmiss} by $\Delta\phi > 0.5$. Further requirements include $\mt > 50\GeV$, $\ptmiss > 55\GeV$, and $\stuv > 300\GeV$, where {\stuv} is defined as the scalar sum of the {\pt} of the two jets matched to leptons ($\ell=\mu,\nu$), the muon, and the {\ptmiss} in the event.

The main backgrounds that can mimic the LQ signal in the {\munujj} channel are $\PW$+jets and {\ttbar}+jets events. Both backgrounds are calculated using simulated samples normalized to the number of events in two separated data control regions. They are estimated with events that, in addition to satisfying the {\munujj} preselection, also satisfy $70 < M_{\mathrm{T}}^{\PGm\Pgn} < 110\GeV$. The events are then separated into two control regions, further enriched in their respective background processes, using \PQb{} tagging. The $\PW$+jets background control region requires no \PQb-tagged jets, while the {\ttbar}+jets control sample requires at least one \PQb-tagged jet. The $\PW$+jets data normalization scale factor is found to be $0.93\pm 0.01\stat$, and the {\ttbar}+jets data normalization scale factor is found to be $0.98\pm 0.01\stat$. As the scale factors do not depend on the kinematic distributions, no further systematic uncertainty is applied. These data normalization scale factors are then applied to simulated events passing the final selections.

Backgrounds from single top quark, $\cPZ/\gamma^*$+jets, and diboson events are estimated from simulation. Background from QCD multijets are shown to be negligible using data control regions.

\begin{figure}[htb]
       \centering
       {\includegraphics[width=.49\textwidth]{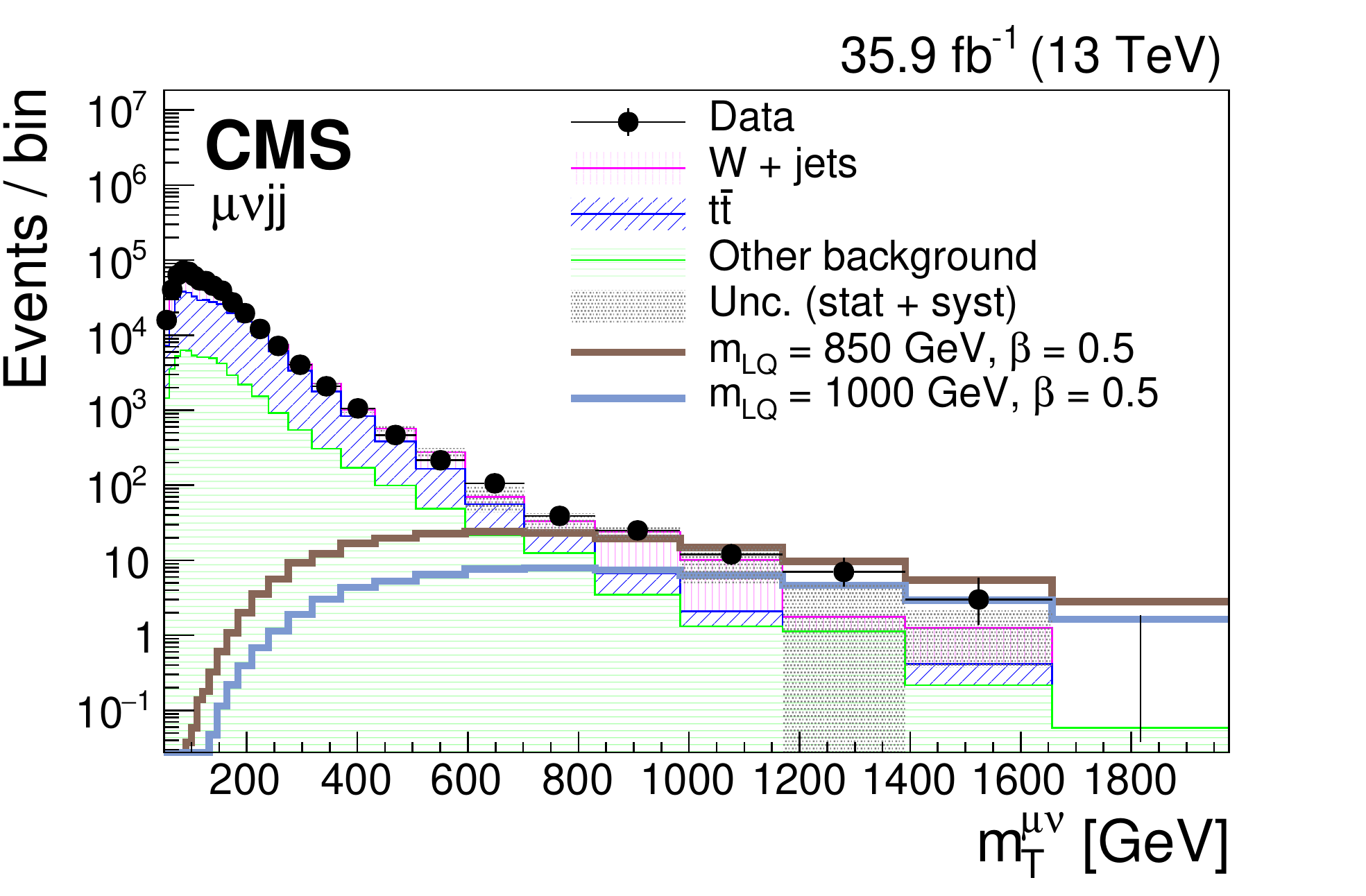}}\\
       {\includegraphics[width=.49\textwidth]{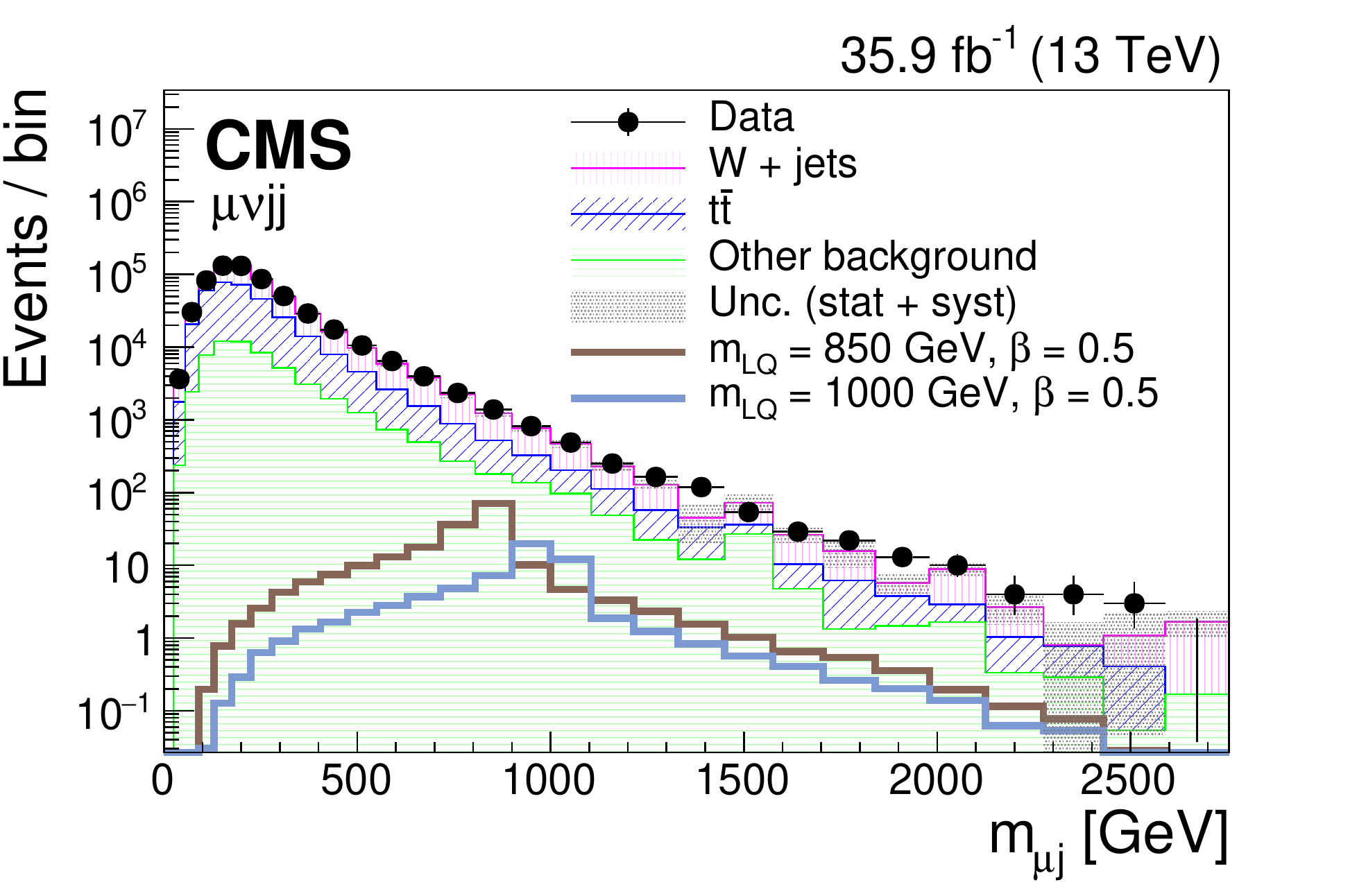}}
       {\includegraphics[width=.49\textwidth]{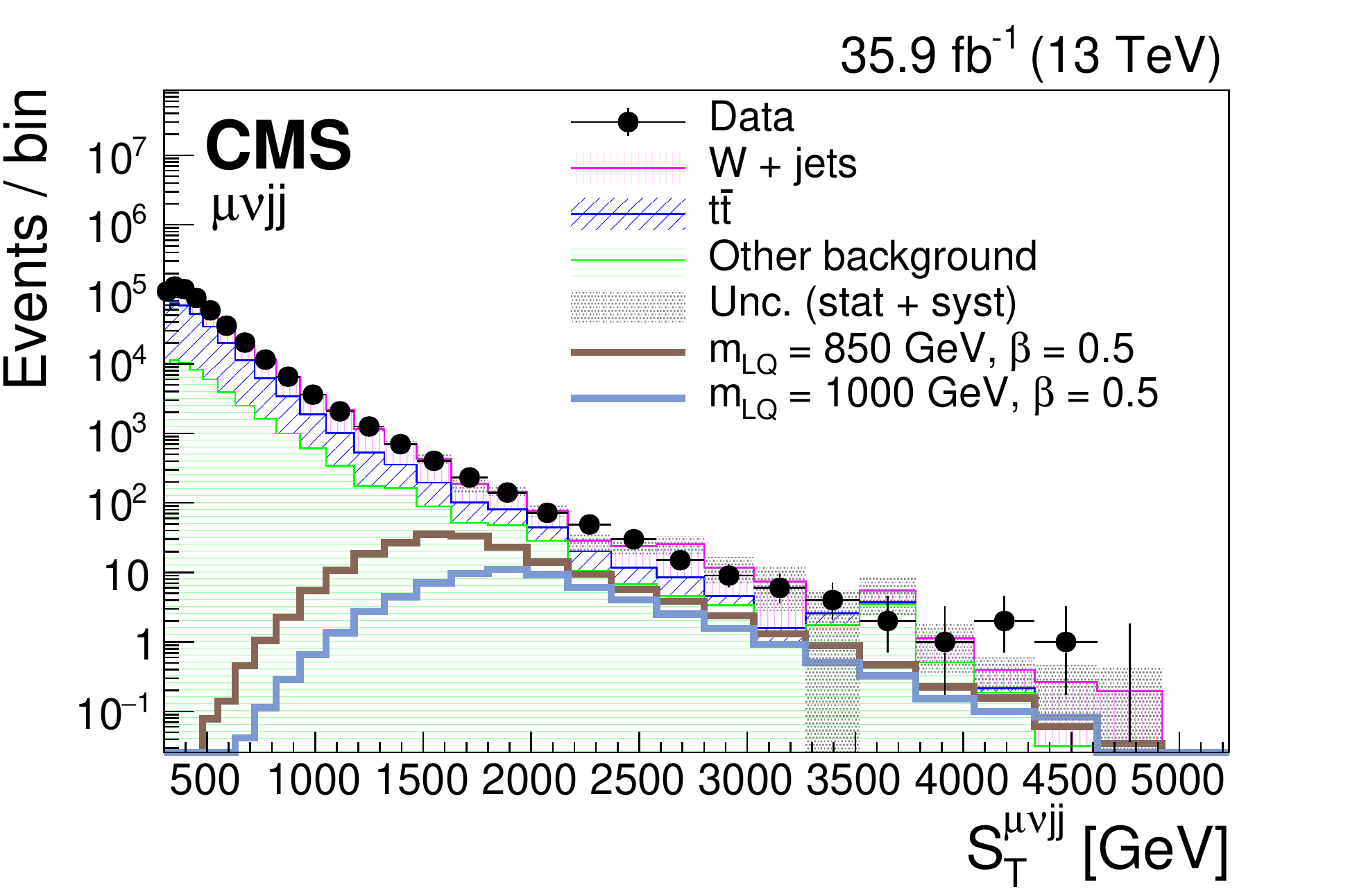}}
       \caption{Comparison of data and background at the preselection level for the {\munujj} channel, for the variables used for final selection criteria optimization: {\mt} (upper), {\Muj} (\cmsLL), and {\stuv} (\cmsLR). `Other background' includes $\cPZ/\gamma^*$+jets, single top quark, and diboson backgrounds. The hashed band represents the combined statistical and systematic uncertainty in the full background estimate.}
          \label{fig:LQMuNu}
\end{figure}

After preselection, discriminating variables are identified, as with the {\mumujj} channel. In the {\munujj} channel, these variables are {\stuv}, {\mt}, and {\Muj}, where {\mt} and {\Muj} are defined as the muon-{\ptvecmiss} transverse mass and the muon-jet invariant mass for the combination that minimizes the LQ-$\overline{\mathrm{LQ}}$ transverse mass difference. Distributions for these variables in events satisfying the preselection are shown in Fig.~\ref{fig:LQMuNu}.

\section{Final selection}
\label{sec:resultsSelection}

\subsection{Final selection optimization}

\begin{figure}[htbp]
       \centering
       {\includegraphics[width=.49\textwidth]{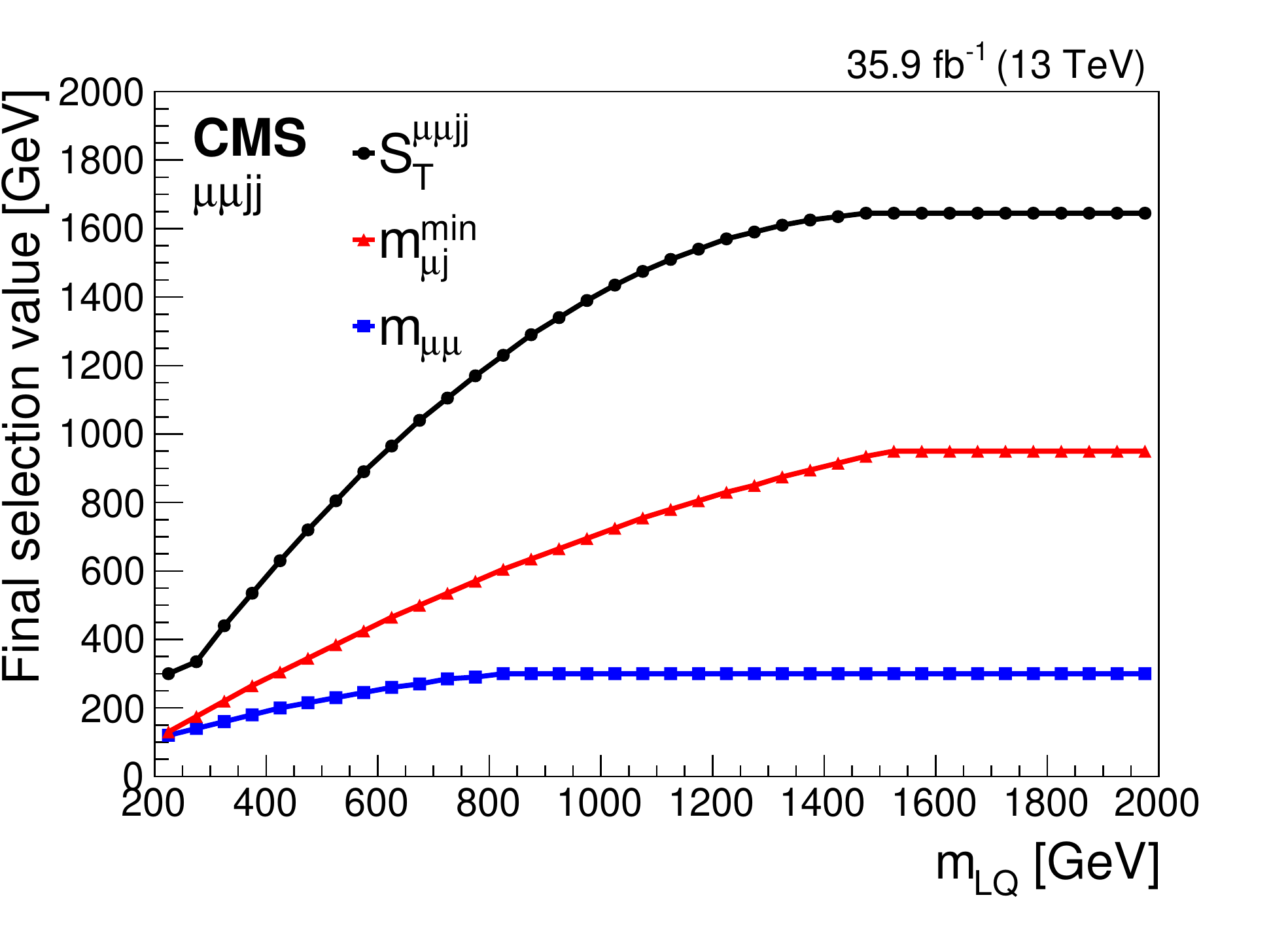}}
       {\includegraphics[width=.49\textwidth]{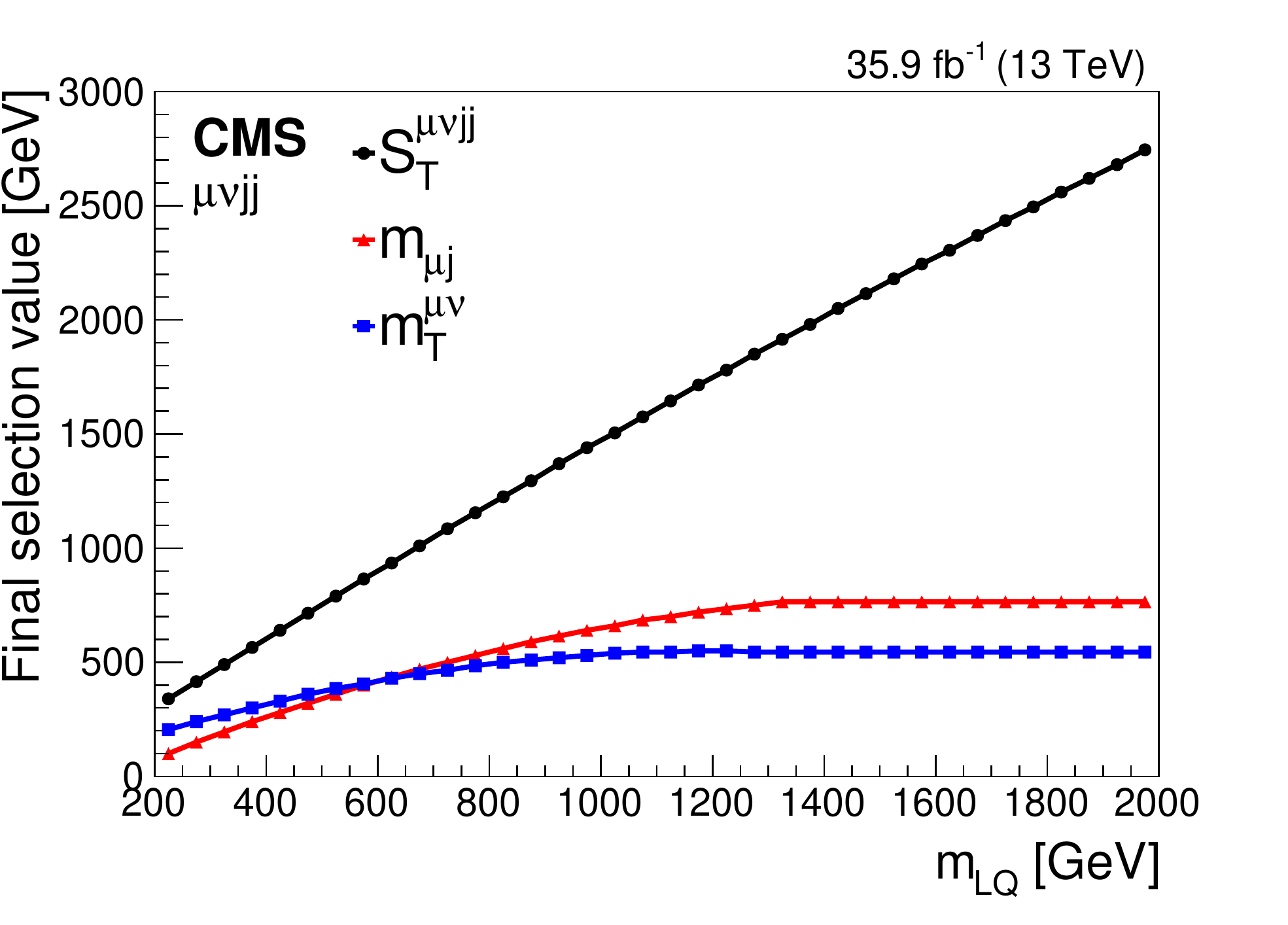}}
       \caption{Lower bounds of the final selection criteria for the three variables for the {\mumujj} (\cmsLeft) and {\munujj} (\cmsRight) channels as a function of scalar {\mlq}.}
          \label{fig:optimization}
\end{figure}

For both the {\mumujj} and {\munujj} channels, the previously described kinematic variables identified as having strong discrimination power between signal and background are used to define a final selection for each {\mlq}. The signal-to-background separation is optimized with a full three-dimensional optimization using the Punzi significance~\cite{Punzi} for a discovery potential of 5 standard deviations at 95\% confidence level (\CL). This method is optimal for both making a discovery and for setting limits, and is valid in cases with low background event counts. In the {\mumujj} channel, the {\Muu} is required to be greater than 100\GeV to exclude the background control region. In the {\munujj} channel, the {\mt} is required to be greater than 110\GeV for the same reason. The lower bounds of the final selection criteria for the three variables are shown as a function of scalar {\mlq} in Fig.~\ref{fig:optimization}. The behavior of the different variable responses to the optimization can be attributed to the shapes of the signal distributions of the different variables, as seen in Figs.~\ref{fig:LQMumu} and~\ref{fig:LQMuNu}.

\subsection{Systematic uncertainties}
\label{sec:syst}

Systematic uncertainties in the LQ signal production cross sections vary from 14 to 50\% across the full LQ mass range.  They are estimated by varying the PDF eigenvectors within their uncertainties and the renormalization and factorization scales by factors of one-half and two.

Systematic uncertainties in the background yields and in the signal acceptance for both the {\mumujj} and {\munujj} channels are calculated for each final selection by running the full analysis with separately varied detector quantities, particle momenta, or scale factors. These yields are compared to those for the nominal analysis, and the differences are propagated as log-normal nuisance parameters in the limit setting. The effects of these systematic uncertainties in signal acceptance and total background yield are shown for the {\mumujj} and {\munujj} channels in Tables~\ref{tab:syst_uujj} and~\ref{tab:syst_uvjj}, respectively.

Systematic uncertainties in the jet energy resolution and muon energy resolution are measured by smearing the jet and muon momenta, including high-{\pt} specific corrections for muons~\cite{Zprime}. Uncertainties due to the jet energy scale and the muon energy scale are estimated by propagating jet and muon energy corrections.

Uncertainties in the shapes of the main backgrounds are estimated by varying the factorization and normalization scales in the simulation by factors of 1/2 and 2. These uncertainties, which include the uncertainty in the extrapolation of the distributions of the final selection variables from the control to signal regions, are estimated for the $\cPZ/\gamma^*$+jets, {\ttbar}+jets, $\PW$+jets, and diboson backgrounds.

In the {\mumujj} channel the uncertainty in the $\cPZ/\gamma^*$+jets background normalization is estimated by varying the normalization scale factor described in Section~\ref{sec:back:mumu} up and down by its statistical and systematic uncertainties added in quadrature. The uncertainty in the {\ttbar}+jets normalization is estimated by varying the {\Pgm\Pgm}/{\Pe\Pgm} correction factor up and down by its statistical uncertainty. In the {\munujj} channel the uncertainties in the $\PW$+jets and {\ttbar}+jets normalizations are estimated by varying the normalization scale factors described in Section~\ref{sec:back:munu} up and down by their statistical uncertainties.

Other sources of systematic uncertainty considered are: the luminosity measurement~\cite{lumi}, muon identification and isolation~\cite{MuId}, pileup~\cite{pileup}, trigger efficiency, and track reconstruction efficiency. The uncertainty from the PDF prediction is estimated by varying the \textsc{NNPDF3.0} eigenvectors within their uncertainties, following the PDF4LHC prescription~\cite{pdf-syst-1,pdf-syst-2}. A further uncertainty in the \PQb{} tagging efficiency is applied only in the {\munujj} channel~\cite{btag}, where the control region is defined via \PQb{} tagging. For most values of {\mlq} the systematic uncertainties are at the lower end of the range. The maximum values given in Tables~\ref{tab:syst_uujj} and~\ref{tab:syst_uvjj} are only relevant for large values of \mlq, where the total uncertainty is dominated by the statistical uncertainty in the simulated background samples.

\begin{table}[hbtp]
\centering
\topcaption{Range of systematic uncertainties in the signal acceptance and background yields for the {\mumujj} analysis. The last two lines show the total systematic uncertainty and the total statistical uncertainty in the simulated samples, respectively.}
\begin{scotch}{lcc}
{\mumujj} uncertainty &  Signal (\%)  &  Background (\%) \\ \hline
Jet energy resolution  &   0.0 -- 0.4   &   0.3 -- 4.8   \\
Jet energy scale       &   0.1 -- 1.8   &   0.4 -- 4.9   \\
Integrated luminosity  &   2.5 -- 2.5   &   0.3 -- 0.9   \\
Muon energy resolution &   0.0 -- 0.2   &   0.0 -- 3.8   \\
Muon energy scale      &   0.0 -- 0.2   &   1.3 -- 6.2   \\
Muon ID/Isolation      &   6.1 -- 6.8   &   1.2 -- 2.9   \\
PDF                    &   1.9 -- 4.0   &   0.4 -- 4.6   \\
Pileup                 &   0.0 -- 0.3   &   0.2 -- 5.9   \\
Trigger                &   0.1 -- 0.7   &   0.0 -- 0.5   \\
Tracking efficiency    &   1.0 -- 2.0   &   0.1 -- 0.9   \\
{\ttbar}+jets normalization    &  \NA   &   0.0 -- 0.3   \\
{\ttbar}+jets  shape           &  \NA   &   0.0 -- 0.0   \\
\PW+jets normalization &  \NA   &   0.0 -- 0.1   \\
\PW+jets shape         &  \NA   &   0.0 -- 0.0   \\
$\cPZ/\gamma^*$+jets normalization        &  \NA   &   3.4 -- 7.3   \\
$\cPZ/\gamma^*$+jets shape                &  \NA   &   1.5 -- 6.2   \\
Diboson shape          &  \NA   &   0.7 -- 9.2           \\ \hline
Total syst. uncertainty       &  7.2 -- 8.5      &   5.0 -- 12  \\
Total stat. uncertainty       &  0.5 -- 1.0        &   0.6 -- 29\\
\end{scotch}
\label{tab:syst_uujj}
\end{table}

\begin{table}[hbtp]
\centering
\topcaption{Range of systematic uncertainties in the signal acceptance and background yields for the {\munujj} analysis. The last two lines show the total systematic uncertainty and the total statistical uncertainty in the simulated samples, respectively.}
\begin{scotch}{lcc}
{\munujj} uncertainty &  Signal (\%)  &  Background (\%) \\ \hline
Jet energy resolution  &   1.2 -- 2.3   &   3.4 -- 6.1   \\
Jet energy scale       &   0.0 -- 0.8   &   0.7 -- 6.7   \\
Integrated Luminosity  &   2.5 -- 2.5   &   0.5 -- 1.4   \\
Muon energy resolution &   0.0 -- 0.1   &   0.2 -- 4.7   \\
Muon energy scale      &   0.0 -- 0.2   &   0.4 -- 2.9   \\
Muon ID/Isolation      &   3.0 -- 3.1   &   0.5 -- 2.5   \\
PDF                    &   0.4 -- 0.8   &   0.9 -- 5.6   \\
Pileup                 &   0.0 -- 0.3   &   0.6 -- 3.1   \\
Trigger                &   4.2 -- 7.5   &   0.8 -- 5.5   \\
Tracking efficiency    &   0.5 -- 1.0   &   0.1 -- 0.7   \\
b tagging efficiency    &   \NA   &   1.4 -- 3.6   \\
{\ttbar}+jets normalization   &   \NA   &   0.1 -- 0.5   \\
{\ttbar}+jets shape           &   \NA   &   0.0 -- 0.0   \\
\PW+jets normalization &   \NA   &   0.3 -- 0.5   \\
\PW+jets shape         &   \NA   &   1.6 -- 8.7   \\
$\cPZ/\gamma^*$+jets normalization        &   \NA   &   0.6 -- 1.4   \\
$\cPZ/\gamma^*$+jets shape                &   \NA   &   0.0 -- 0.0   \\
Diboson shape          &   \NA   &   0.5 -- 8.4           \\ \hline
Total syst. uncertainty       &  6.1 -- 8.7      &   6.6 -- 13 \\
Total stat. uncertainty       &  0.1 -- 1.3      &   0.2 -- 19 \\
\end{scotch}
\label{tab:syst_uvjj}
\end{table}

\section{Results}
\label{sec:results}

\subsection{Data comparison with background after final selection}

\begin{figure*}[htb]
       \centering
       {\includegraphics[width=.49\textwidth]{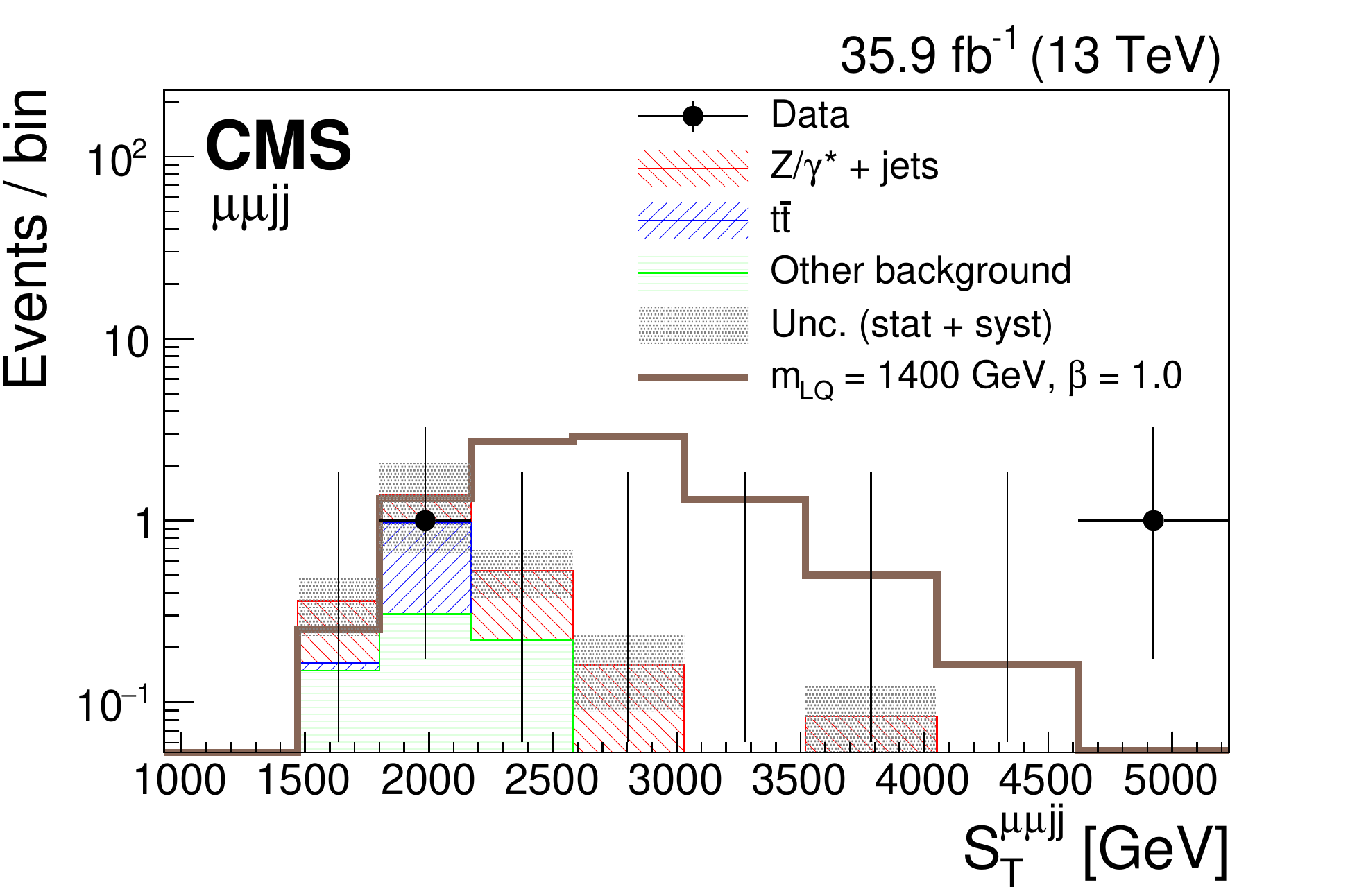}}
       {\includegraphics[width=.49\textwidth]{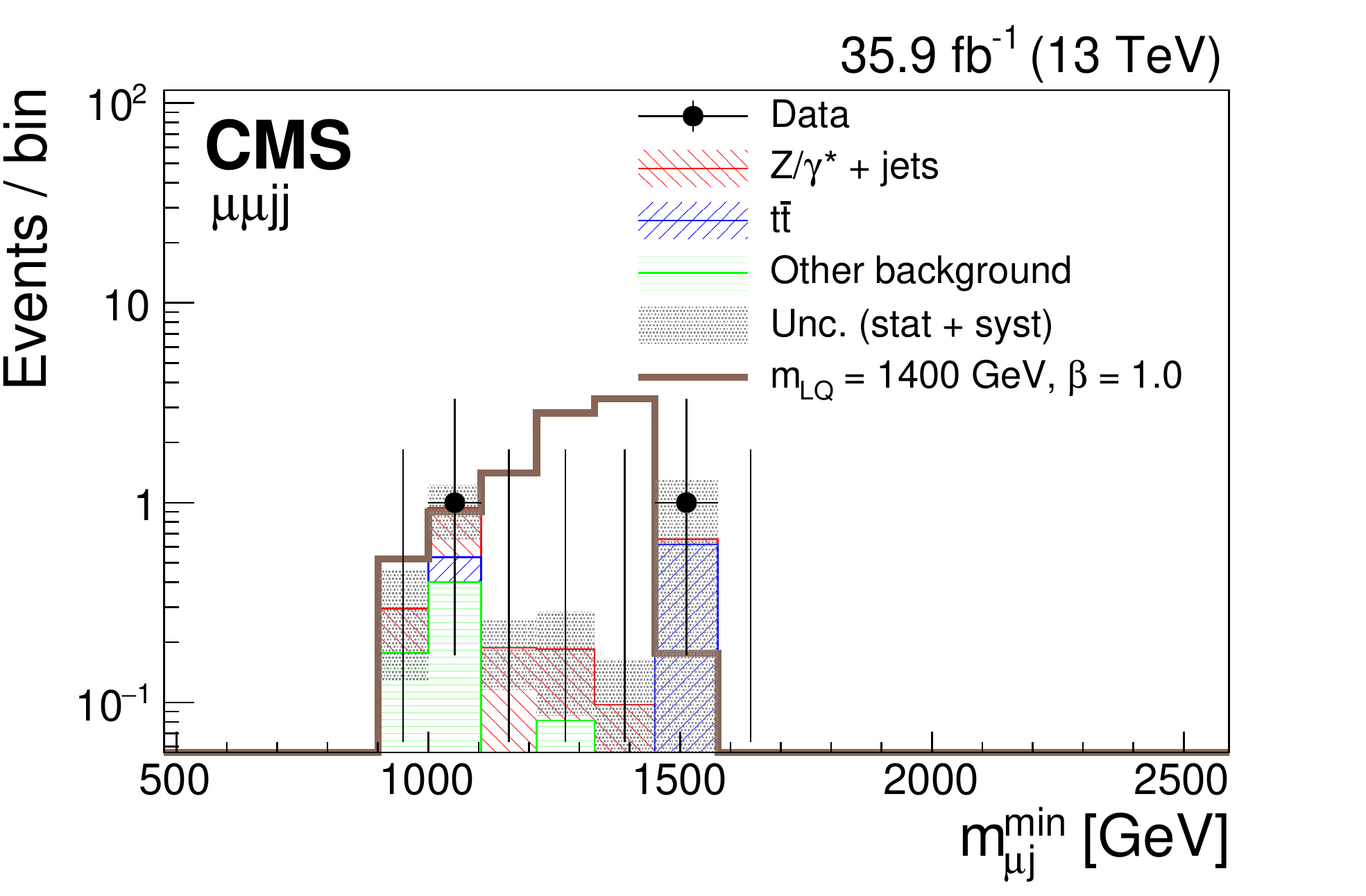}}
       {\includegraphics[width=.49\textwidth]{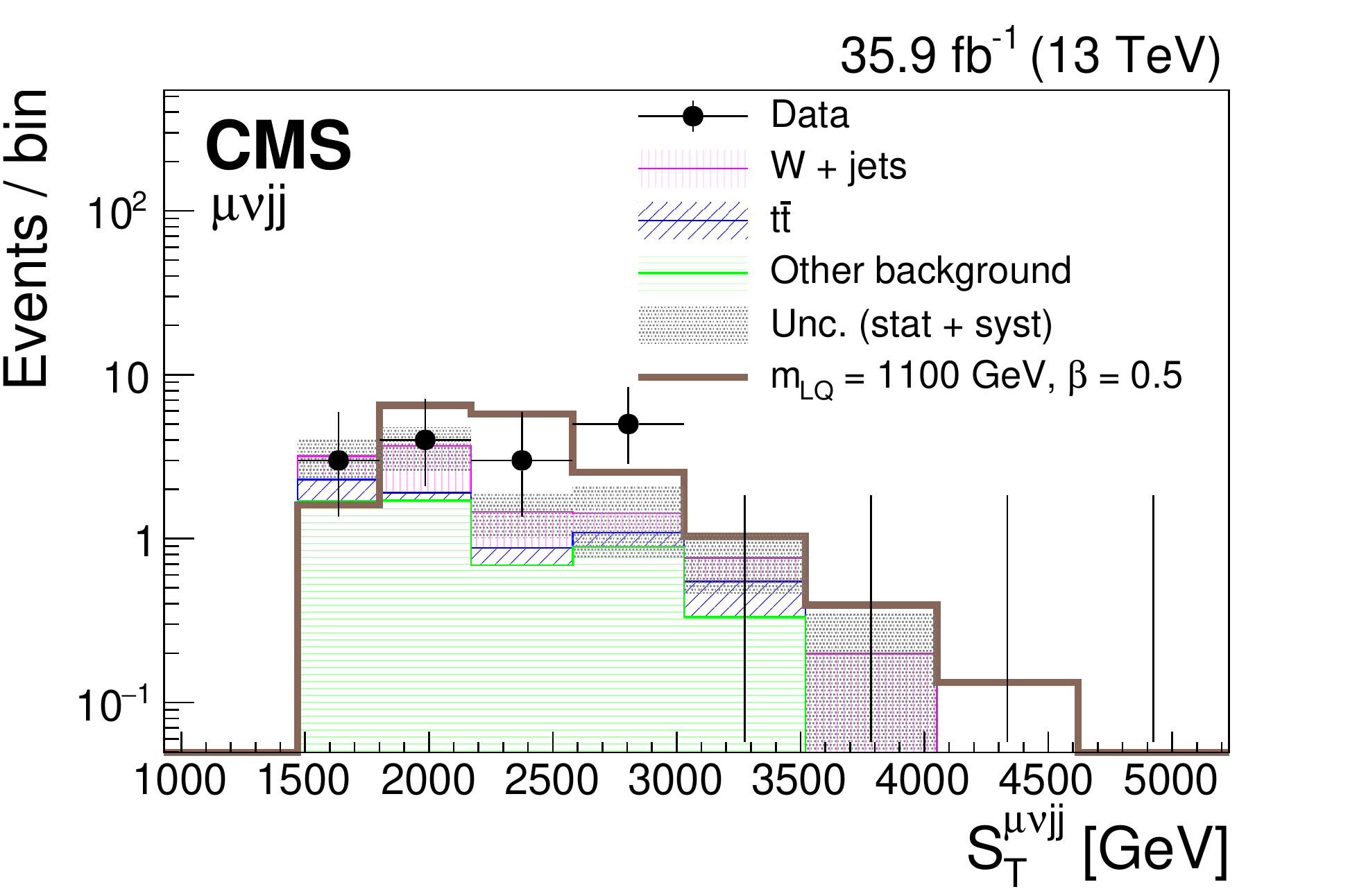}}
       {\includegraphics[width=.49\textwidth]{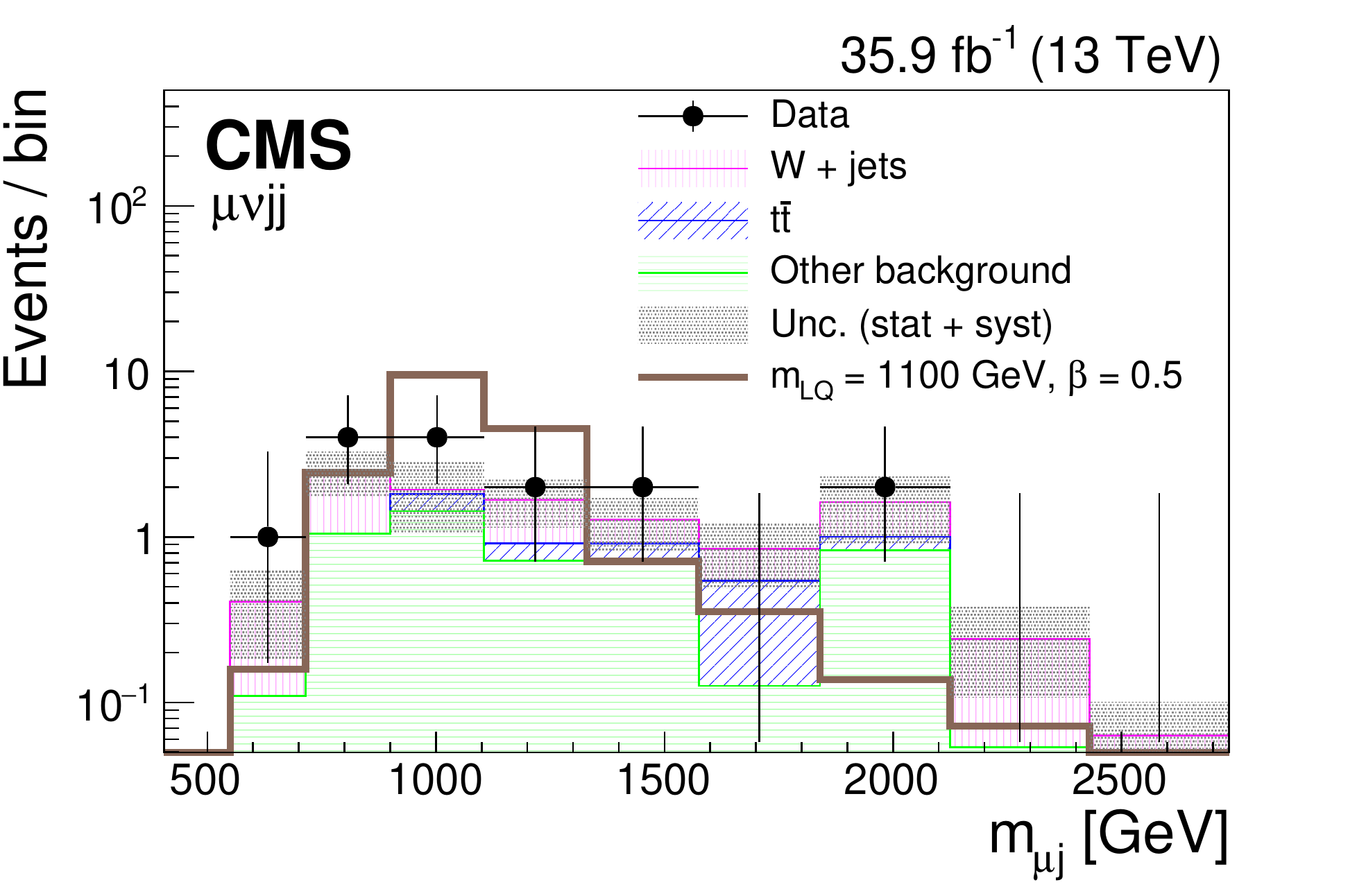}}
       \caption{Comparison of data and background distributions of {\st} (left), {\Mujmin} (upper right), and {\Muj} (lower right), for the {\mumujj} channel (upper plots) and the {\munujj} channel (lower plots). Events after final selections with $\mlq = 1400\GeV$ are shown in the upper plots, and with $\mlq = 1100\GeV$ in the lower plots. The hashed band represents the combined statistical and systematic uncertainty in the full background estimate. `Other background' includes $\PW$+jets, single top quark, and diboson backgrounds in the {\mumujj} channel, and $\cPZ/\gamma^*$+jets, single top quark, and diboson backgrounds in the {\munujj} channel.}
          \label{fig:finalSel}
\end{figure*}

The data are compared to background predictions after the final selections have been applied. Comparisons of the kinematic distributions, after the final selection, for data and simulation for two {\mlq} hypotheses are shown in Fig.~\ref{fig:finalSel}. No significant excess above the predicted background is seen for any {\mlq}, within uncertainties. The largest difference between data and the background estimate is a roughly two standard deviation excess in the {\munujj} channel for $\mlq = 950\GeV$. Kinematic distributions of the small excess of events in this region do not look like signal events, lacking the characteristic mass peak expected of LQs. There is one high-{\st} event that can be seen in Fig.~\ref{fig:finalSel} (upper left) that merits mention. The background estimate for high mass final selections for $\st > 3000$\GeV is $0.0^{+0.1}_{-0.0}$. However, this event is unlike a signal event. In particular, the invariant masses of the two LQ candidates in this event are not compatible with LQ pair production.

\begin{figure}[htbp]
       \centering
       {\includegraphics[width=\cmsFigWidth]{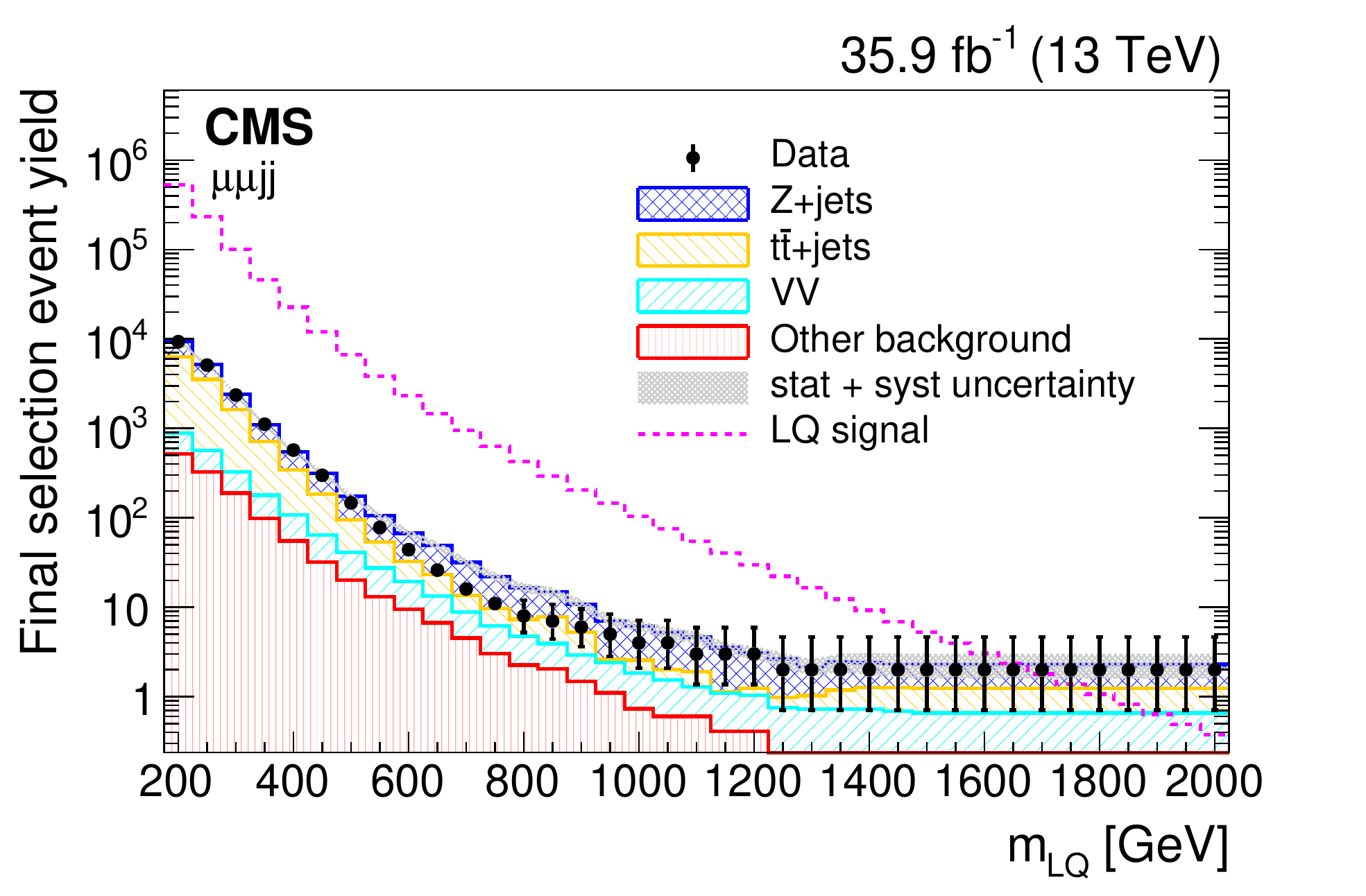}}
       \caption{Data and background event yields after final selections for the {\mumujj} analysis, as a function of scalar \mlq. `Other background' includes $\PW$+jets and single top quark production. The selection criteria for each bin are detailed in Table~\ref{tab:syst_uujj}. All the bins are correlated, as the events selected for each {\mlq} are a strict subset of the events selected for the lower mass LQ. The hashed band represents the combined statistical and systematic uncertainty in the full background estimate.}
          \label{fig:finalSelTableuujj}
\end{figure}

\begin{figure}[htbp]
       \centering
       {\includegraphics[width=\cmsFigWidth]{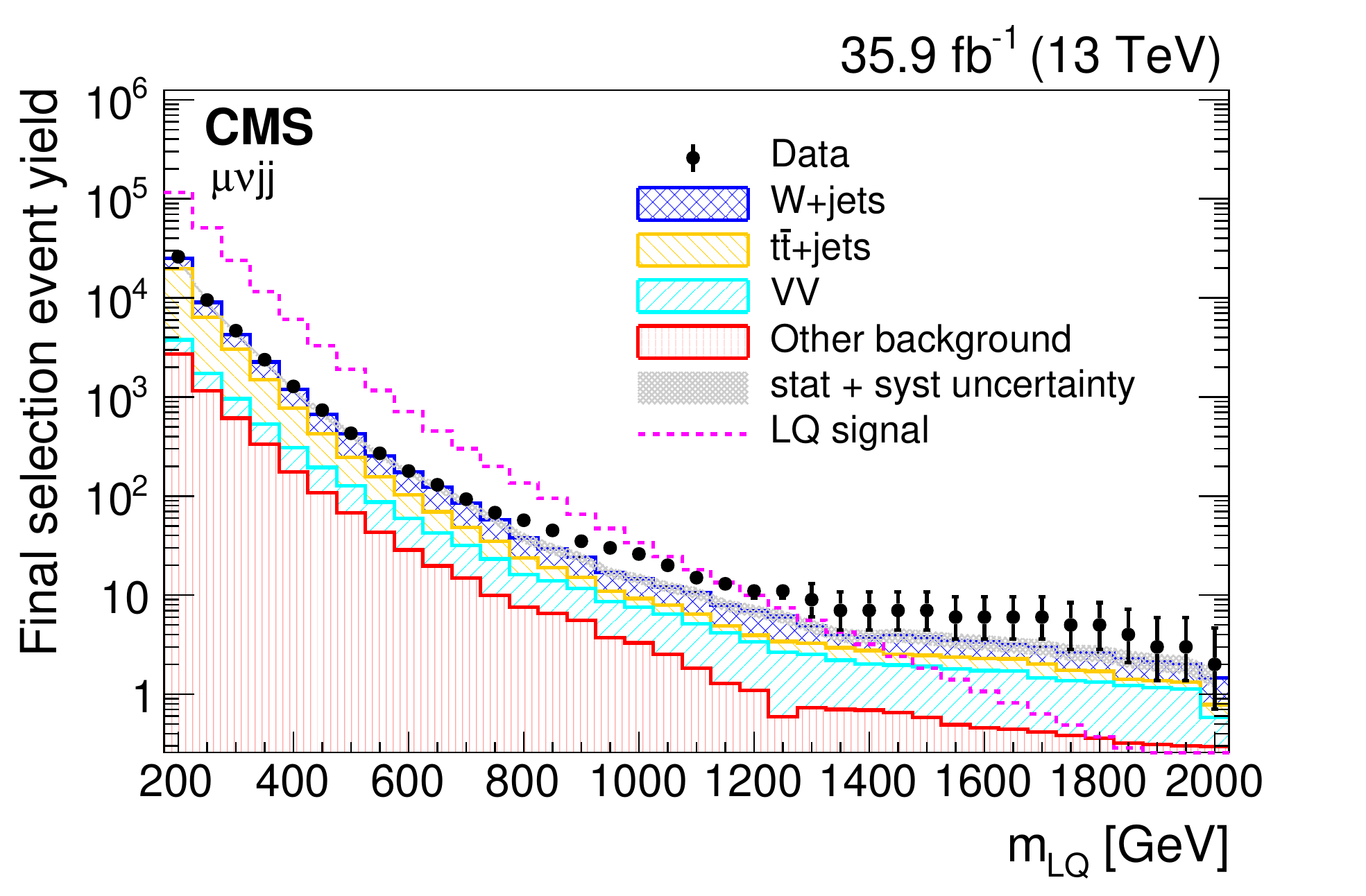}}
       \caption{Data and background event yields after final selections for the {\munujj} analysis, as a function of \mlq. `Other background' includes $\cPZ/\gamma^*$+jets and single top quark production. The selection criteria for each bin are detailed in Table~\ref{tab:syst_uvjj}. All the bins are correlated, as the events selected for each {\mlq} are a strict subset of the events selected for the lower mass LQ. The hashed band represents the combined statistical and systematic uncertainty in the full background estimate.}
          \label{fig:finalSelTableuvjj}
\end{figure}

Comparisons of background, data, and signal for each set of final selections can be seen in Figs.~\ref{fig:finalSelTableuujj} and~\ref{fig:finalSelTableuvjj}. The $y$ axis shows the final selection event yields for each of the individual {\mlq} hypotheses shown on the $x$ axis. All the bins are correlated in these plots, as the events selected for each {\mlq} are a strict subset of the events selected for the lower mass LQ. The product of acceptance and efficiency of the signal for all final selections, as well as detailed tables of the event counts in data, background, and signal, are shown in Appendix~\ref{appendix}.

\subsection{Limit setting}
Limits are set on the LQ pair production cross section $\sigma$ as a function of scalar {\mlq}, using the asymptotic approximation~\cite{Cowan:2010js} of the modified frequentist {\CLs} approach~\cite{cls1,cls2}, utilizing the ratio of the confidences in the signal+background to background hypotheses. The systematic uncertainties listed above are introduced as nuisance parameters in the limit setting procedure using log-normal probability functions. Uncertainties of a statistical nature are described by $\Gamma$ distributions with widths determined by the number of events in simulated samples or observed in data control regions. These limits have been compared to the so-called `LHC-style' fully-frequentist {\CLs} limits~\cite{cls3} and are found to be in good agreement with the expected and observed limits for all final selections, but with slightly more conservative systematic uncertainties in the low background regime.

\begin{figure}[htbp!]
  \centering
  \includegraphics[width=0.48\textwidth]{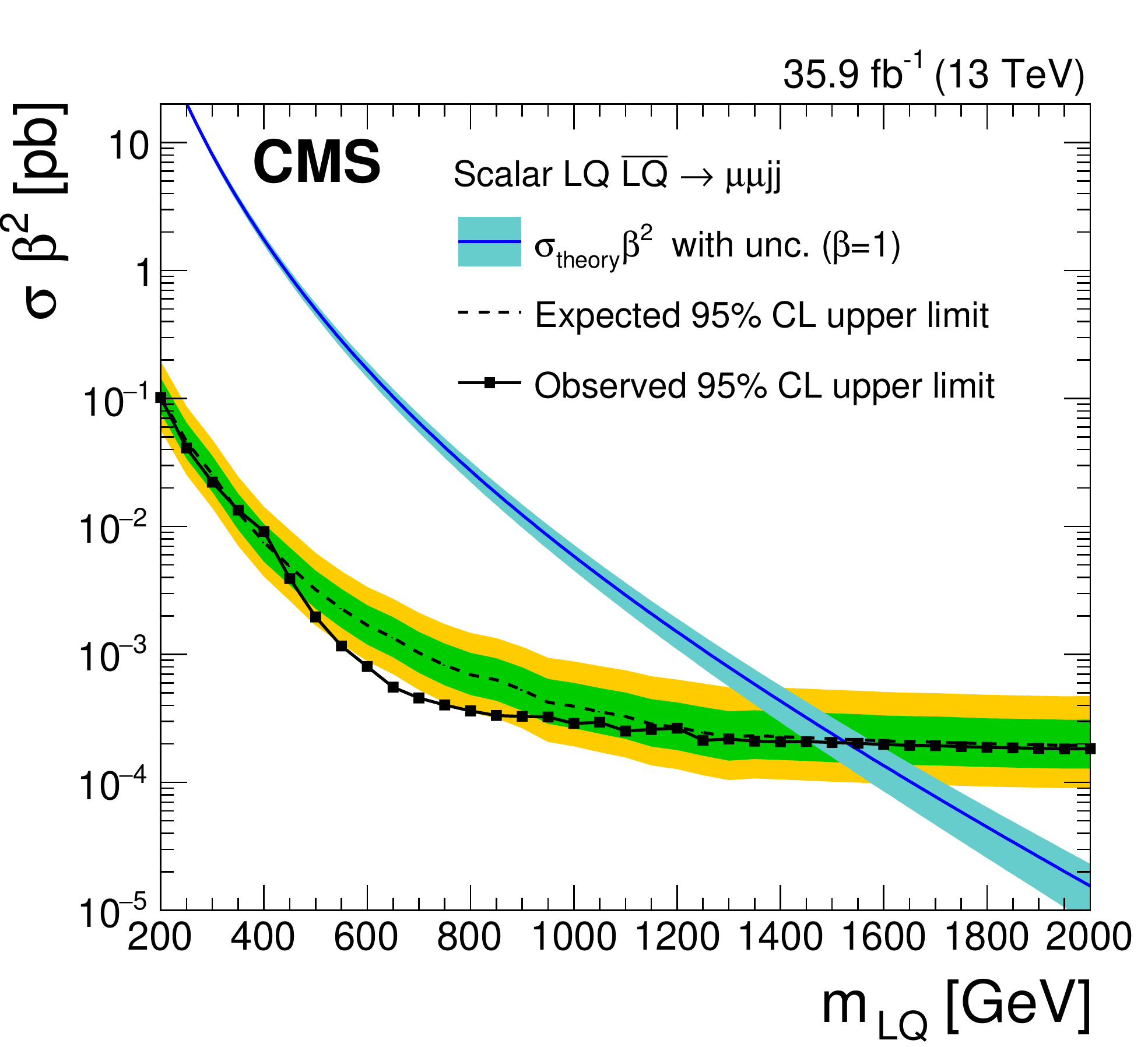}
  \includegraphics[width=0.48\textwidth]{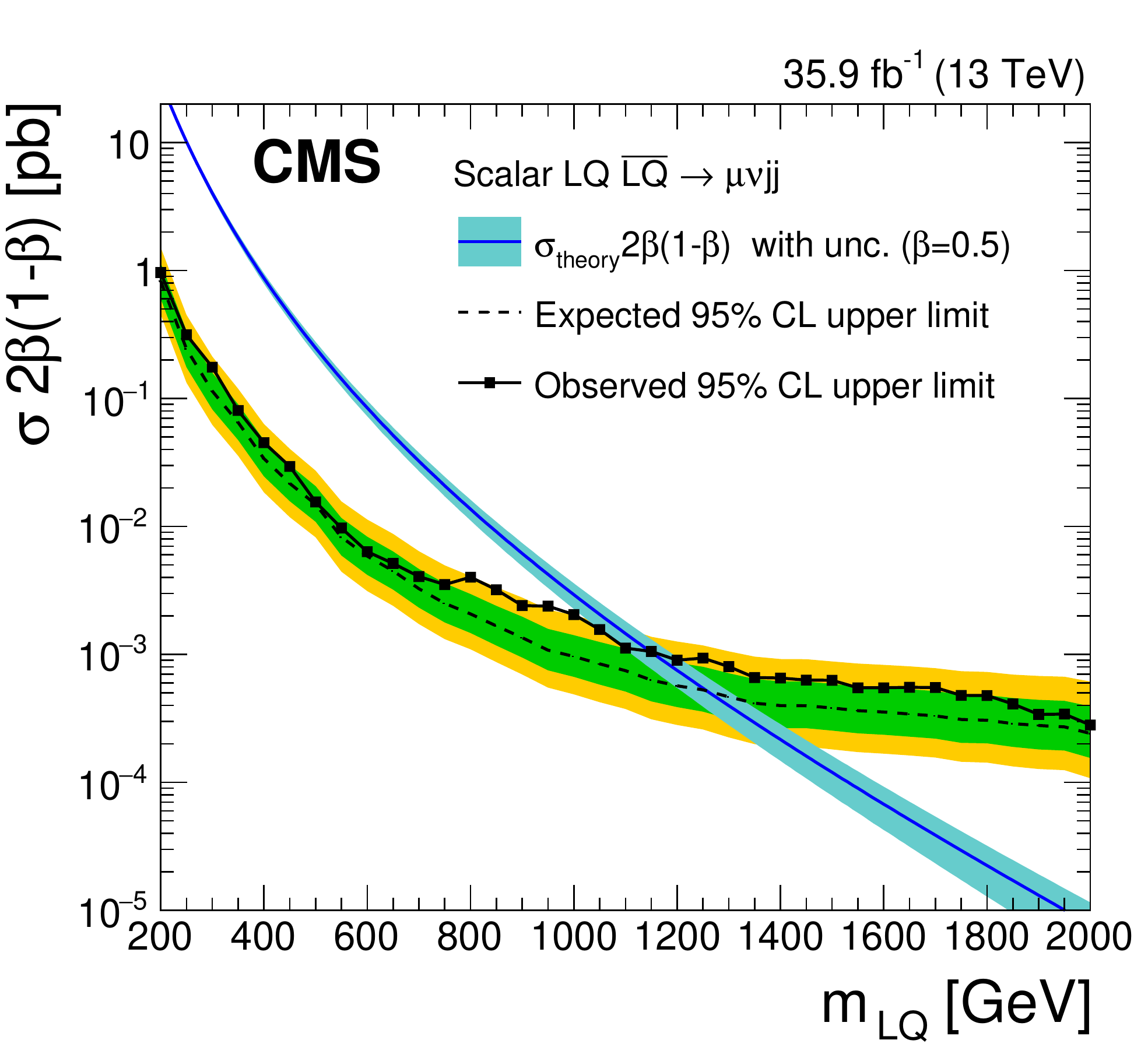}
   \caption{The expected and observed upper limits at 95\% {\CL} on the product of the scalar LQ pair production cross section and the branching fractions $\beta^2$ or $2\beta(1-\beta)$ as a function of the second-generation {\mlq} obtained with the {\mumujj} (\cmsLeft) or {\munujj} (\cmsRight) analysis. The solid lines represent the observed limits, the dashed lines represent the median expected limits, and the inner dark-green and outer light-yellow bands represent the 68\% and 95\% confidence intervals. The $\sigma_{\text{theory}}$ curves and their blue bands represent the theoretical scalar LQ pair production cross sections and the uncertainties on the cross sections due to the PDF prediction and renormalization and factorization scales, respectively.
    }
  \label{fig:limit_plots_mu}
\end{figure}

The 95\% {\CL} upper limits on $\sigma\,\beta^2$ or $\sigma\,2\beta(1-\beta)$ as a function of scalar {\mlq} are shown, together with the NLO predictions for the scalar LQ pair production cross section, in Fig.~\ref{fig:limit_plots_mu}. Systematic uncertainties in the LQ signal production cross sections are shown as a band around the signal production cross section. By comparing the observed upper limit with the theoretical cross section values, second-generation scalar LQs with masses less than 1530 (1150)\GeV are excluded under the assumption that $\beta = 1.0$\,(0.5), compared to the median expected limits of 1515\,(1260)\GeV.

\begin{figure}[htbp!]
  \centering
  \includegraphics[width=0.48\textwidth]{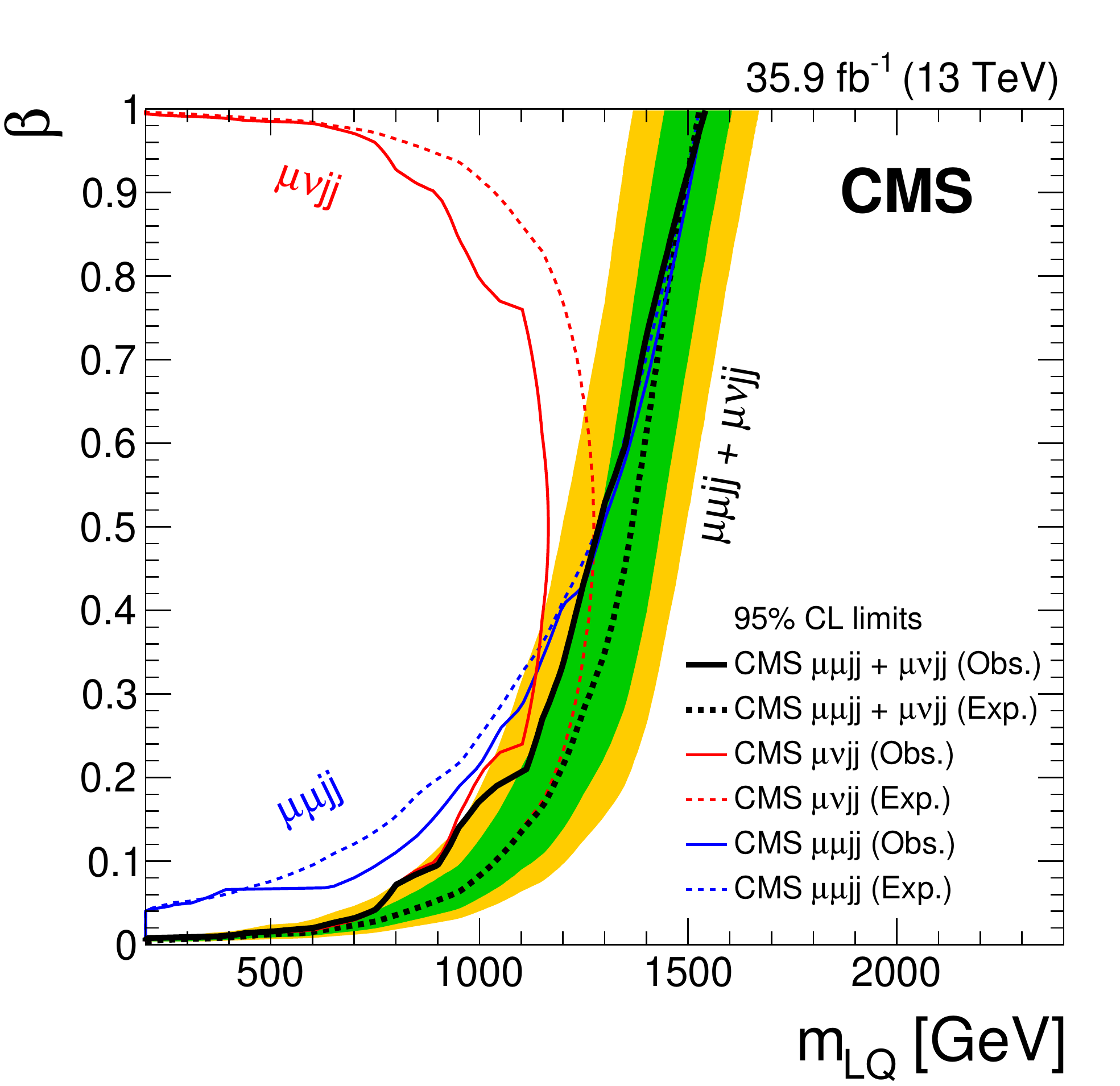}
   \caption{The expected and observed exclusion limits at 95\% {\CL} for second-generation {\mlq} as a function of the branching fraction $\beta$ vs. \mlq. The inner dark-green and outer light-yellow expected limit uncertainty bands represent the 68\% and 95\% confidence intervals on the combination. Limits for the individual {\mumujj} and {\munujj} channels are also drawn. The solid lines represent the observed limits in each channel, and the dashed lines represent the expected limits.}
  \label{fig:limit_plots_lqcombo}
\end{figure}

Limits are also set at 95\% {\CL} for $\beta$ values from 0 to 1 for both the {\mumujj} and {\munujj} channels, as well as for the combination of both channels. In the combination, all systematic uncertainties are treated as fully correlated and all statistical uncertainties are treated as fully uncorrelated. The resulting two-dimensional limit plot is shown in Fig.~\ref{fig:limit_plots_lqcombo}. The combination of the two channels improves the mass exclusion, particularly for low values of $\beta$. Using the combined channels, second-generation scalar LQs with masses less than 1285\GeV can be excluded for $\beta = 0.5$, compared with an expected limit of 1365\GeV.

\begin{figure}[bht!]
  \centering
  \includegraphics[width=0.48\textwidth]{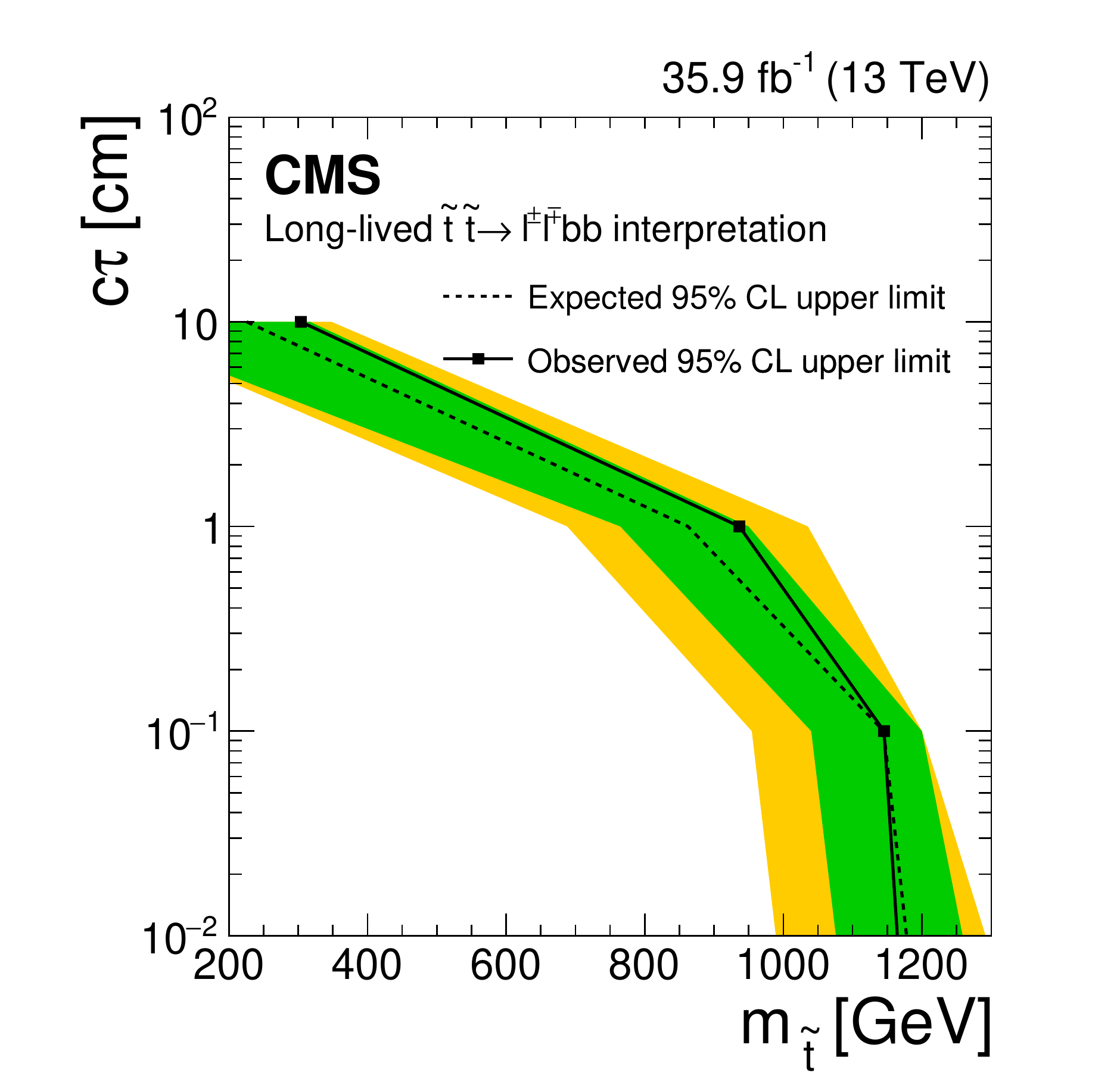}
  \caption{Expected and observed upper limits at 95\% {\CL} on the long-lived RPV SUSY {\PSQt} pair production cross section as a function of {\PSQt} mass ($x$ axis) and lifetime ($y$ axis). The dashed line and the inner dark-green and outer light-yellow uncertainty bands represent the median expected limits, and the 68\% and 95\% confidence intervals, respectively.}
  \label{fig:longlivedlimit}
\end{figure}

The results in the {\mumujj} channel are also interpreted in the context of the displaced SUSY model described in Section~\ref{sec:introduction}. Studies in both simulation and data have shown that tracking efficiency remains at $\sim$100\% for the lifetimes and corresponding impact parameters considered~\cite{Khachatryan:2014mea}, which allows interpretation of the results for a displaced signal to be made with the same final selections and systematic uncertainties as previously used for a prompt signal. The 95\% {\CL} expected and observed limits on the displaced SUSY {\PSQt} pair production cross section are shown in Fig.~\ref{fig:longlivedlimit}. The limits are presented in two dimensions as a function of {\PSQt} mass and lifetime.  The expected and observed limits have been extrapolated to $c\tau = 0\unit{cm}$ using the prompt LQ limits, taking into account the assumed {\PSQt} branching ratio, $\PSQt\to \mathrm{b} \mu = 1/3$. This extrapolation connects these results to the prompt kinematic range and is motivated by the fact that prompt top squark pair production is kinematically very similar to that for LQs. The observed exclusion limits are 1150, 940, and 305\GeV for $c\tau = 0.1$, 1.0, and 10.0\unit{cm}. Following the formulation in Ref.~\cite{Richardson:2000nt} these limits can be translated into lower bounds on the coupling strength of the RPV term in the SUSY Lagrangian, in this case $\lambda'_{233}\mathrm{cos}(\theta)$, where $\mathrm{cos}(\theta)$ represents the mixing angle between the left- and right-handed eigenstates of the top squarks. Using the expression for the partial width $\Gamma(\PSQt\to \mathrm{b}\ell)=3\Gamma(\PSQt\to \mathrm{b}\mu)\approx 3c (\lambda'_{233}\mathrm{cos}(\theta))^2m_{\PSQt}/16\pi$~\cite{Richardson:2000nt}, the excluded regions correspond to $\lambda'_{233}\mathrm{cos}(\theta) < 5.4\times 10^{-8}$, $1.9\times 10^{-8}$, and $1.0\times 10^{-8}$, respectively, for the mass and lifetime limits described above. These limits provide complementary sensitivity to dedicated searches for long-lived particles~\cite{Khachatryan:2014mea}, which generally require particles with longer decay lengths in their triggers.

\section{Summary}
\label{sec:summary}
A search has been presented for pair production of second-generation leptoquarks using proton-proton collision data collected at $\sqrt{s} = 13\TeV$ in 2016 with the CMS detector at the LHC, corresponding to an integrated luminosity of 35.9\fbinv. Limits are set at 95\% confidence level on the product of the scalar leptoquark pair production cross section and $\beta^2$ ($2\beta(1-\beta)$) in the {\mumujj} (\munujj) channels, as a function of the leptoquark mass \mlq.  Second-generation leptoquarks with masses less than 1530\,(1285)\GeV are excluded for $\beta=1.0$\,(0.5), an improvement of 370 (525)\GeV compared to previously published results. Two-dimensional limits are set in the $\beta$--{\mlq} plane. The results in the {\mumujj} search are interpreted in the context of an $R$-parity violating supersymmetry model with long-lived top squarks. These limits represent the most stringent limits to date on these models.

\clearpage
\begin{acknowledgments}
We congratulate our colleagues in the CERN accelerator departments for the excellent performance of the LHC and thank the technical and administrative staffs at CERN and at other CMS institutes for their contributions to the success of the CMS effort. In addition, we gratefully acknowledge the computing centers and personnel of the Worldwide LHC Computing Grid for delivering so effectively the computing infrastructure essential to our analyses. Finally, we acknowledge the enduring support for the construction and operation of the LHC and the CMS detector provided by the following funding agencies: BMWFW and FWF (Austria); FNRS and FWO (Belgium); CNPq, CAPES, FAPERJ, and FAPESP (Brazil); MES (Bulgaria); CERN; CAS, MoST, and NSFC (China); COLCIENCIAS (Colombia); MSES and CSF (Croatia); RPF (Cyprus); SENESCYT (Ecuador); MoER, ERC IUT, and ERDF (Estonia); Academy of Finland, MEC, and HIP (Finland); CEA and CNRS/IN2P3 (France); BMBF, DFG, and HGF (Germany); GSRT (Greece); NKFIA (Hungary); DAE and DST (India); IPM (Iran); SFI (Ireland); INFN (Italy); MSIP and NRF (Republic of Korea); LAS (Lithuania); MOE and UM (Malaysia); BUAP, CINVESTAV, CONACYT, LNS, SEP, and UASLP-FAI (Mexico); MBIE (New Zealand); PAEC (Pakistan); MSHE and NSC (Poland); FCT (Portugal); JINR (Dubna); MON, RosAtom, RAS and RFBR (Russia); MESTD (Serbia); SEIDI, CPAN, PCTI and FEDER (Spain); Swiss Funding Agencies (Switzerland); MST (Taipei); ThEPCenter, IPST, STAR, and NSTDA (Thailand); TUBITAK and TAEK (Turkey); NASU and SFFR (Ukraine); STFC (United Kingdom); DOE and NSF (USA).

\hyphenation{Rachada-pisek} Individuals have received support from the Marie-Curie program and the European Research Council and Horizon 2020 Grant, contract No. 675440 (European Union); the Leventis Foundation; the A. P. Sloan Foundation; the Alexander von Humboldt Foundation; the Belgian Federal Science Policy Office; the Fonds pour la Formation \`a la Recherche dans l'Industrie et dans l'Agriculture (FRIA-Belgium); the Agentschap voor Innovatie door Wetenschap en Technologie (IWT-Belgium); the F.R.S.-FNRS and FWO (Belgium) under the ``Excellence of Science - EOS" - be.h project n. 30820817; the Ministry of Education, Youth and Sports (MEYS) of the Czech Republic; the Lend\"ulet (``Momentum") Programme and the J\'anos Bolyai Research Scholarship of the Hungarian Academy of Sciences, the New National Excellence Program \'UNKP, the NKFIA research grants 123842, 123959, 124845, 124850 and 125105 (Hungary); the Council of Science and Industrial Research, India; the HOMING PLUS program of the Foundation for Polish Science, cofinanced from European Union, Regional Development Fund, the Mobility Plus program of the Ministry of Science and Higher Education, the National Science Center (Poland), contracts Harmonia 2014/14/M/ST2/00428, Opus 2014/13/B/ST2/02543, 2014/15/B/ST2/03998, and 2015/19/B/ST2/02861, Sonata-bis 2012/07/E/ST2/01406; the National Priorities Research Program by Qatar National Research Fund; the Programa Estatal de Fomento de la Investigaci{\'o}n Cient{\'i}fica y T{\'e}cnica de Excelencia Mar\'{\i}a de Maeztu, grant MDM-2015-0509 and the Programa Severo Ochoa del Principado de Asturias; the Thalis and Aristeia programs cofinanced by EU-ESF and the Greek NSRF; the Rachadapisek Sompot Fund for Postdoctoral Fellowship, Chulalongkorn University and the Chulalongkorn Academic into Its 2nd Century Project Advancement Project (Thailand); the Welch Foundation, contract C-1845; and the Weston Havens Foundation (USA).
\end{acknowledgments}

\bibliography{auto_generated}
\appendix
\section{Efficiencies and event yields}
\label{appendix}

\begin{figure*}[htb]
 \centering
 {\includegraphics[width=.49\textwidth]{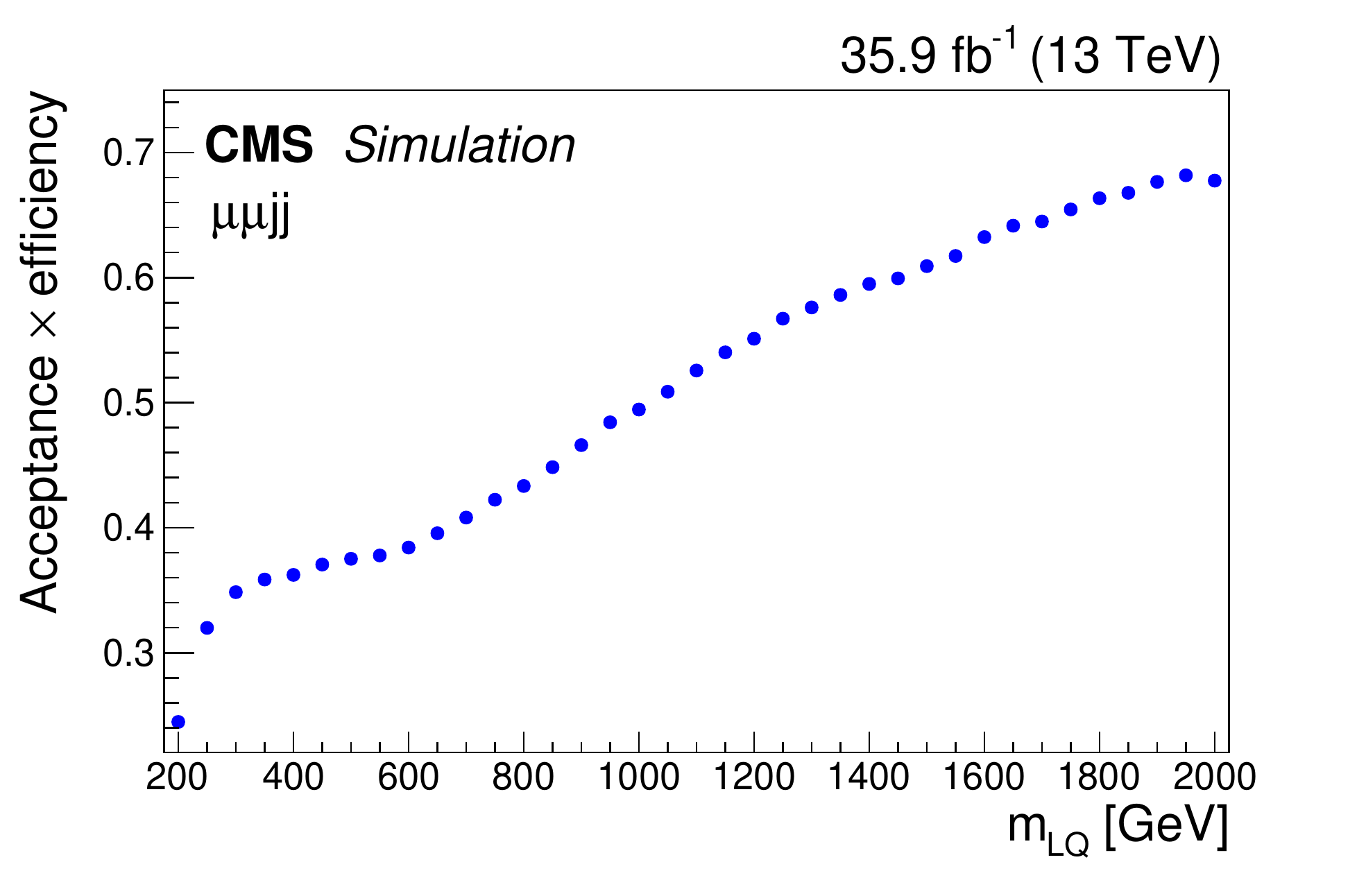}}
 {\includegraphics[width=.49\textwidth]{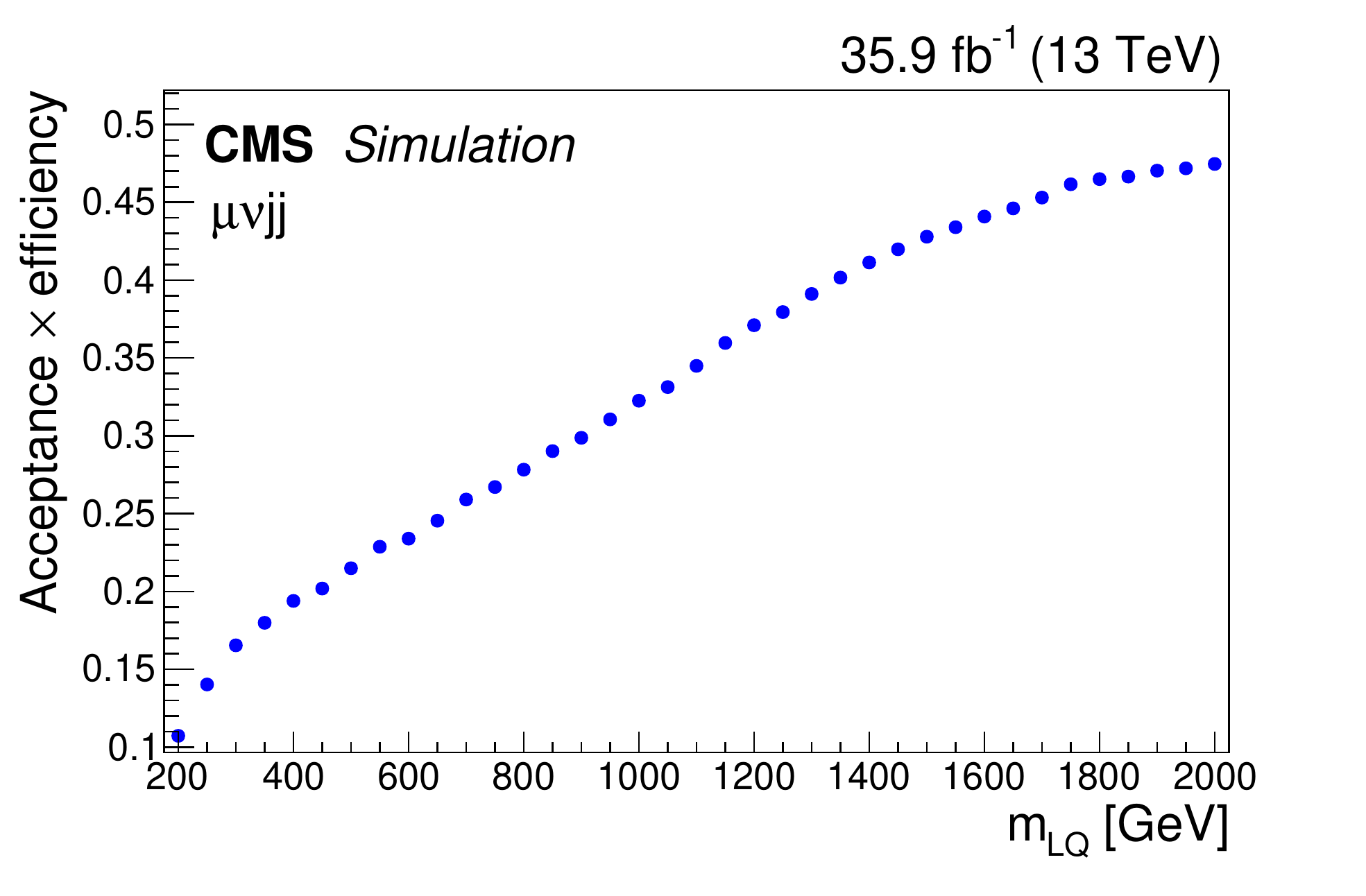}}
 \caption{The product of signal acceptance and efficiency for optimized final selections as a function of {\mlq} in the {\mumujj} (left) and {\munujj} (right) channels.}
 \label{fig:signalEff}
\end{figure*}

The product of signal acceptance and efficiency for optimized final selections as a function of {\mlq} in the {\mumujj} (left) and {\munujj} (right) channels is shown in Fig.~\ref{fig:signalEff}. Event yields at final selection level for the {\mumujj} and {\munujj} analyses are shown in Tables~\ref{tab:finalselection_mumu} and~\ref{tab:finalselection_munu}, respectively.

\begin{table*}[htbp]
\centering
\topcaption{Event yields after final selections for the {\mumujj} analysis. `Other bkg.' includes $\PW$+jets and single top quark production. Uncertainties are statistical unless otherwise indicated.}
\cmsTable{
\begin{scotch}{ccccccccc}
{\mlq} ({\GeVns}) &	 Signal &              	 $\cPZ/\gamma^*$+jets &            	 {\ttbar}+jets &          	 Diboson &                	 Other bkg. &                     	  All bkg. (stat + syst)&                  	 Data \\ \hline	
200 &     	 531700 $\pm$ 4700  &  	 2973 $\pm$ 7  & 	 5467 $\pm$ 56  &    	 369 $\pm$ 2  &    	 519 $\pm$ 10  &             	 9328 $\pm$ 57  $\pm$ 444  &             9317 \\
\vspace{0.13cm}
250 &     	 232900 $\pm$ 1800  &  	 1675 $\pm$ 5  &   	 2972 $\pm$ 41  &    	 241 $\pm$ 2  &  	 324 $\pm$ 8  &             	 5213 $\pm$ 42  $\pm$ 250  &             5102 \\
\vspace{0.13cm}
300 &     	 100460 $\pm$ 760  &   	 793 $\pm$ 3  &  	 1298 $\pm$ 26  &    	 138 $\pm$ 1  &  	 189 $\pm$ 6  &             	 2419 $\pm$ 27  $\pm$ 117  &             2360 \\
\vspace{0.13cm}
350 &     	 46160 $\pm$ 340  &    	 3878 $\pm$ 2  &  	 538 $\pm$ 16  &     	 81.0 $\pm$ 1.0  &   	 98.0 $\pm$ 4.1  &               1105 $\pm$ 17  $\pm$ 57  &              1113 \\
\vspace{0.13cm}
400 &     	 22610 $\pm$ 160  &    	 202 $\pm$ 1  &  	 237 $\pm$ 10  &     	 51.9 $\pm$ 0.8  & 	 55.2 $\pm$ 3.1  &               546 $\pm$ 11  $\pm$ 29  &               572 \\
\vspace{0.13cm}
450 &     	 12039 $\pm$ 86  &     	 132 $\pm$ 1  &  	 121 $\pm$ 7  &  	 32.2 $\pm$ 0.7  & 	 31.8 $\pm$ 2.3  &               316 $\pm$ 78  $\pm$ 18  &          	 299 \\
\vspace{0.13cm}
500 &     	 6672 $\pm$ 48  &      	 79.0 $\pm$ 0.7  &    	 54.1 $\pm$ 4.6  &   	 20.9 $\pm$ 0.5  & 	 20.2 $\pm$ 1.9  &               174 $\pm$ 5  $\pm$ 11  &          	 147 \\
\vspace{0.13cm}
550 &     	 3848 $\pm$ 27  &      	 52.0 $\pm$ 0.5  &  	 26.1 $\pm$ 3.0  &   	 14.4 $\pm$ 0.5  & 	 13.1 $\pm$ 1.5  &               106 $\pm$ 3  $\pm$ 8  &           	 78 \\
\vspace{0.13cm}
600 &     	 2328 $\pm$ 16  &      	 34.7 $\pm$ 0.4  & 	 12.9 $\pm$ 1.9  &   	 10.0 $\pm$ 0.3  & 	 9.44 $\pm$ 1.27  &             	 67.0 $\pm$ 2.4  $\pm$ 5.2  &            44 \\
\vspace{0.13cm}
650 &     	 1461 $\pm$ 10  &      	 26.0 $\pm$ 0.3  & 	 9.90 $\pm$ 1.80  &    	 6.55 $\pm$ 0.30  &   	 6.70 $\pm$ 1.10  &               49.0 $\pm$ 2.1  $\pm$ 3.9  &            26 \\
\vspace{0.13cm}
700 &     	 948 $\pm$ 7  &    	 18.2 $\pm$ 0.3  & 	 4.68 $\pm$ 1.07  &  	 4.36 $\pm$ 0.24  &  	 4.53 $\pm$ 0.91  &             	 32.0 $\pm$ 1.4  $\pm$ 2.6  &            16 \\
\vspace{0.13cm}
750 &     	 630 $\pm$ 4  &    	 12.4 $\pm$ 0.2  & 	 3.47 $\pm$ 0.93  &  	 3.17 $\pm$ 0.20  &   	 3.04 $\pm$ 0.74  &             	 22.0 $\pm$ 1.2  $\pm$ 1.9  &            11 \\
\vspace{0.13cm}
800 &     	 424 $\pm$ 3  &    	 9.18 $\pm$ 0.16  &  	 2.62 $\pm$ 0.83  &  	 2.45 $\pm$ 0.19  &  	 2.26 $\pm$ 0.63  &             	 16.5 $\pm$ 1.1  $\pm$ 1.6  &            8 \\
\vspace{0.13cm}
850 &     	 293 $\pm$ 2  &    	 6.93 $\pm$ 0.13  &  	 3.89 $\pm$ 1.23  &  	 1.88 $\pm$ 0.17  &  	 2.05 $\pm$ 0.60  &               14.8 $\pm$ 1.4  $\pm$ 1.1  &            7 \\
\vspace{0.13cm}
900 &     	 206 $\pm$ 1  &    	 5.55 $\pm$ 0.11  &  	 2.34 $\pm$ 0.88  &  	 1.44 $\pm$ 0.15  &  	 1.49 $\pm$ 0.50  &               10.8 $\pm$ 1.0  $\pm$ 0.9  &         	 6 \\
\vspace{0.13cm}
950 &     	 147 $\pm$ 1  &  	 4.41 $\pm$ 0.10  &	 0.22 $\pm$ 0.13  &  	 1.31 $\pm$ 0.15  &  	 1.11 $\pm$ 0.43  &           	 7.04 $\pm$ 0.48  $\pm$ 0.71  &          5 \\
\vspace{0.13cm}
1000 &    	 103.9 $\pm$ 0.7  &  	 3.66 $\pm$ 0.09  &	 0.72 $\pm$ 0.42  &  	 1.10 $\pm$ 0.13  &   	 0.73 $\pm$ 0.33  &           	 6.21 $\pm$ 0.56  $\pm$ 0.59  &          4 \\
\vspace{0.13cm}
1050 &    	 75.0 $\pm$ 0.5  &   	 3.23 $\pm$ 0.09  &	 0.47 $\pm$ 0.33  & 	 0.93 $\pm$ 0.12  &  	 0.60 $\pm$ 0.31  &           	 5.24 $\pm$ 0.48  $\pm$ 0.56  &          4 \\
\vspace{0.13cm}
1100 &    	 54.9 $\pm$ 0.3  &   	 2.71 $\pm$ 0.07  &	 0.60 $\pm$ 0.43  &	 0.69 $\pm$ 0.10  &   	 0.60 $\pm$ 0.31  &           	 4.60 $\pm$ 0.54  $\pm$ 0.48  &          3 \\
\vspace{0.13cm}
1150 &    	 40.3 $\pm$ 0.2  &    	 2.39 $\pm$ 0.07  & 	 0.04 $\pm$ 0.04  &	 0.69 $\pm$ 0.10  &   	 0.41 $\pm$ 0.25  &           	 3.53 $\pm$ 0.28  $\pm$ 0.42  &          3 \\
\vspace{0.13cm}
1200 &    	 29.7 $\pm$ 0.2  &   	 1.86 $\pm$ 0.06  &	 0.19 $\pm$ 0.19  &	 0.63 $\pm$ 0.10  &   	 0.41 $\pm$ 0.25  &           	 3.10 $\pm$ 0.33  $\pm$ 0.42  &          3 \\
\vspace{0.13cm}
1250 &    	 22.2 $\pm$ 0.1  &   	 1.68 $\pm$ 0.06  &	 0.22 $\pm$ 0.22  &	 0.56 $\pm$ 0.10  &	 0.20 $\pm$ 0.19  &           	 2.65 $\pm$ 0.31  $\pm$ 0.34  &          2 \\
\vspace{0.13cm}
1300 &    	 16.4 $\pm$ 0.1  & 	 1.13 $\pm$ 0.04  & 	 0.30 $\pm$ 0.30  &	 0.53 $\pm$ 0.10  &   	 0.12 $\pm$ 0.19  &           	 2.15 $\pm$ 0.37  $\pm$ 0.27  &          2 \\
\vspace{0.13cm}
1350 &    	 12.3 $\pm$ 0.1  &  	 1.26 $\pm$ 0.05  &	 0.46 $\pm$ 0.46  &  	 0.53 $\pm$ 0.10  &   	 0.20 $\pm$ 0.19  &           	 2.45 $\pm$ 0.51  $\pm$ 0.24  &          2 \\
\vspace{0.13cm}
1400 &    	 9.24 $\pm$ 0.05  &  	 1.14 $\pm$ 0.04  &	 0.54 $\pm$ 0.54  &	 0.54 $\pm$ 0.11  &  	 0.19 $ _{-0.19}^{+0.28}$   &	 2.41 $ _{-0.59}^{+0.62}$   $\pm$ 0.24  &	 2 \\
\vspace{0.13cm}
1450 &    	 6.90 $\pm$ 0.04  &  	 1.06 $\pm$ 0.04  &	 0.58 $\pm$ 0.58  &	 0.50 $\pm$ 0.11  &   	 0.19 $ _{-0.19}^{+0.28}$   &	 2.32 $ _{-0.62}^{+0.65}$   $\pm$ 0.22  &	 2 \\
\vspace{0.13cm}
1500 &    	 5.24 $\pm$ 0.03  &  	 1.05 $\pm$ 0.05  &	 0.59 $\pm$ 0.59  &	 0.47 $\pm$ 0.11  &  	 0.19 $ _{-0.19}^{+0.28}$   &	 2.30 $ _{-0.63}^{+0.66}$   $\pm$ 0.23  & 	 2 \\
\vspace{0.13cm}
1550 &    	 3.99 $\pm$ 0.02  &  	 1.05 $\pm$ 0.05  &	 0.59 $\pm$ 0.59  &	 0.47 $\pm$ 0.11  &  	 0.19 $ _{-0.19}^{+0.28}$   &	 2.30 $ _{-0.63}^{+0.66}$   $\pm$ 0.23  & 	 2 \\
\vspace{0.13cm}
1600 &    	 3.06 $\pm$ 0.02  &  	 1.05 $\pm$ 0.05  &	 0.59 $\pm$ 0.59  &	 0.47 $\pm$ 0.11  &  	 0.19 $ _{-0.19}^{+0.28}$   &	 2.30 $ _{-0.63}^{+0.66}$   $\pm$ 0.23  & 	 2 \\
\vspace{0.13cm}
1650 &    	 2.35 $\pm$ 0.01  &  	 1.05 $\pm$ 0.05  &	 0.59 $\pm$ 0.59  &	 0.47 $\pm$ 0.11  &  	 0.19 $ _{-0.19}^{+0.28}$   &	 2.30 $ _{-0.63}^{+0.66}$   $\pm$ 0.23  & 	 2 \\
\vspace{0.13cm}
1700 &    	 1.79 $\pm$ 0.01  &	 1.05 $\pm$ 0.05  &	 0.59 $\pm$ 0.59  &	 0.47 $\pm$ 0.11  &  	 0.19 $ _{-0.19}^{+0.28}$   &	 2.30 $ _{-0.63}^{+0.66}$   $\pm$ 0.23  & 	 2 \\
\vspace{0.13cm}
1750 &    	 1.38 $\pm$ 0.01  &	 1.05 $\pm$ 0.05  &	 0.59 $\pm$ 0.59  &	 0.47 $\pm$ 0.11  &  	 0.19 $ _{-0.19}^{+0.28}$   &	 2.30 $ _{-0.63}^{+0.66}$   $\pm$ 0.23  & 	 2 \\
\vspace{0.13cm}
1800 &    	 1.07 $\pm$ 0.01  &	 1.05 $\pm$ 0.05  &	 0.59 $\pm$ 0.59  &	 0.47 $\pm$ 0.11  &  	 0.19 $ _{-0.19}^{+0.28}$   &	 2.30 $ _{-0.63}^{+0.66}$   $\pm$ 0.23  & 	 2 \\
\vspace{0.13cm}
1850 &    	 0.821 $\pm$ 0.004  &  	 1.05 $\pm$ 0.05  &	 0.59 $\pm$ 0.59  &	 0.47 $\pm$ 0.11  &  	 0.19 $ _{-0.19}^{+0.28}$   &	 2.30 $ _{-0.63}^{+0.66}$   $\pm$ 0.23  & 	 2 \\
\vspace{0.13cm}
1900 &    	 0.636 $\pm$ 0.003  &  	 1.05 $\pm$ 0.05  &	 0.59 $\pm$ 0.59  &	 0.47 $\pm$ 0.11  &  	 0.19 $ _{-0.19}^{+0.28}$   &	 2.30 $ _{-0.63}^{+0.66}$   $\pm$ 0.23  & 	 2 \\
\vspace{0.13cm}
1950 &    	 0.491 $\pm$ 0.003  &  	 1.05 $\pm$ 0.05  &	 0.59 $\pm$ 0.59  &	 0.47 $\pm$ 0.11  &  	 0.19 $ _{-0.19}^{+0.28}$   &	 2.30 $ _{-0.63}^{+0.66}$   $\pm$ 0.23  & 	 2 \\
\vspace{0.13cm}
2000 &    	 0.377 $\pm$ 0.002  &  	 1.05 $\pm$ 0.05  &	 0.59 $\pm$ 0.59  &	 0.47 $\pm$ 0.11  &  	 0.19 $ _{-0.19}^{+0.28}$   &	 2.30 $ _{-0.63}^{+0.66}$   $\pm$ 0.23  & 	 2 \\
\end{scotch}
}
\label{tab:finalselection_mumu}
\end{table*}

\begin{table*}[hbtp]
\centering
\topcaption{Event yields after final selections for the {\munujj} analysis.  `Other bkg.' includes $\cPZ/\gamma^*$+jets and single top quark production. Uncertainties are statistical unless otherwise indicated.}
\cmsTable{
\begin{scotch}{ccccccccc}
{\mlq} ({\GeVns}) &	 Signal &              	 \PW+jets &            	 {\ttbar}+jets &          	 Diboson &                	 Other bkg. &                     	  All bkg. (stat + syst)&                  	 Data \\ \hline	
200 &     	 116600 $\pm$ 1500  &  	 5672 $\pm$ 26  &    	 15816 $\pm$ 51  &   	 1049 $\pm$ 5  & 	 2732 $\pm$ 15  &                25270 $\pm$ 59  $\pm$ 1171  &           26043 \\      	

\vspace{0.13cm}
250 &     	 51050 $\pm$ 580  &    	 2635 $\pm$ 16  &    	 4662 $\pm$ 28  &    	 575 $\pm$ 3  &  	 1155 $\pm$ 10  &                9029 $\pm$ 34  $\pm$ 431  &             9519 \\
\vspace{0.13cm}
300 &     	 23840 $\pm$ 250  &    	 1259 $\pm$ 10  & 	 2066 $\pm$ 18  &    	 346 $\pm$ 3  &  	 611.7 $\pm$ 7  &             	 4284 $\pm$ 22  $\pm$ 197  &             4669 \\
\vspace{0.13cm}
350 &     	 11580 $\pm$ 120  &    	 757 $\pm$ 7  &  	 964 $\pm$ 13  &     	 200 $\pm$ 2  &  	 335 $\pm$ 5  &               	 2256 $\pm$ 16  $\pm$ 122  &             2379 \\
\vspace{0.13cm}
400 &     	 6051 $\pm$ 58  &      	 418 $\pm$ 5  &  	 461 $\pm$ 9  &  	 131 $\pm$ 2  &  	 176 $\pm$ 4  &               	 1187 $\pm$ 11  $\pm$ 70  &              1279 \\
\vspace{0.13cm}
450 &     	 3280 $\pm$ 32  &      	 248 $\pm$ 3  &  	 228 $\pm$ 6  &  	 86.4 $\pm$ 1.6  &   	 108 $\pm$ 3  &             	 671 $\pm$ 8  $\pm$ 47  &              	 737 \\
\vspace{0.13cm}
500 &     	 1911 $\pm$ 18  &      	 177 $\pm$ 3  &  	 119 $\pm$ 4  &  	 58.8 $\pm$ 1.3  &   	 67.6 $\pm$ 2.7  &               422 $\pm$ 6  $\pm$ 40  &            	 430 \\
\vspace{0.13cm}
550 &     	 1165 $\pm$ 10  &      	 99.2 $\pm$ 1.8  &   	 69.2 $\pm$ 3.4  &   	 44.0 $\pm$ 1.2  &     	 42.9 $\pm$ 2.1  &               255 $\pm$ 4  $\pm$ 19  &          	 270 \\
\vspace{0.13cm}
600 &     	 7089 $\pm$ 6  &    	 70.9 $\pm$ 1.5  &   	 43.4 $\pm$ 2.7  &   	 31.1 $\pm$ 1.0  &   	 28.6 $\pm$ 1.7  &               174 $\pm$ 3  $\pm$ 13  &              	 179 \\
\vspace{0.13cm}
650 &     	 453 $\pm$ 4  &    	 53.8 $\pm$ 1.3  &   	 26.8 $\pm$ 2.1  &   	 22.9 $\pm$ 0.91  & 	 19.7 $\pm$ 1.4  &               123 $\pm$ 3  $\pm$ 10  &          	 130 \\
\vspace{0.13cm}
700 &     	 301 $\pm$ 3  &      	 36.0 $\pm$ 1.9  & 	 16.7 $\pm$ 1.7  &   	 17.0 $\pm$ 0.78  & 	 14.8 $\pm$ 1.2  &               84.6 $\pm$ 2.4  $\pm$ 7.1  &            93 \\
\vspace{0.13cm}
750 &     	 199 $\pm$ 2  &    	 22.7 $\pm$ 0.7  & 	 11.6 $\pm$ 1.4  & 	 13.3 $\pm$ 0.71  & 	 9.89 $\pm$ 0.96  &              57.5 $\pm$ 2.0  $\pm$ 5.2  &            68 \\
\vspace{0.13cm}
800 &     	 136 $\pm$ 1  &    	 14.0 $\pm$ 0.5  & 	 7.60 $\pm$ 1.15  &   	 8.58 $\pm$ 0.52  &  	 7.60 $\pm$ 0.83  &              37.7 $\pm$ 1.6  $\pm$ 4.3  &            57 \\
\vspace{0.13cm}
850 &     	 94.7 $\pm$ 0.8  &   	 10.5 $\pm$ 0.4  & 	 4.88 $\pm$ 0.92  &  	 7.46 $\pm$ 0.52  &  	 6.51 $\pm$ 0.81  &              29.3 $\pm$ 1.4  $\pm$ 3.5  &            45 \\
\vspace{0.13cm}
900 &     	 65.9 $\pm$ 0.5  &   	 8.96 $\pm$ 0.34  &  	 3.43 $\pm$ 0.79  &  	 6.14 $\pm$ 0.48  &  	 5.56 $\pm$ 0.75  &              24.1 $\pm$ 1.2  $\pm$ 2.4  &            35 \\
\vspace{0.13cm}
950 &     	 47.1 $\pm$ 0.4  &   	 5.96 $\pm$ 0.25  &  	 2.36 $\pm$ 0.65  &  	 4.85 $\pm$ 0.42  &  	 3.70 $\pm$ 0.55  &              16.9 $\pm$ 1.0  $\pm$ 1.7  &         	 30 \\
\vspace{0.13cm}
1000 &    	 33.9 $\pm$ 0.3  &   	 5.40 $\pm$ 0.24  &   	 1.66 $\pm$ 0.55  &  	 4.30 $\pm$ 0.41  &   	 3.30 $\pm$ 0.52  &              14.7 $\pm$ 0.9  $\pm$ 1.5  &          	 26 \\
\vspace{0.13cm}
1050 &    	 24.4 $\pm$ 0.2  &   	 4.20 $\pm$ 0.20  &    	 1.48 $\pm$ 0.52  &  	 3.90 $\pm$ 0.40  &    	 2.54 $\pm$ 0.45  &              12.1 $\pm$ 0.8  $\pm$ 1.3  &         	 20 \\
\vspace{0.13cm}
1100 &    	 18.0 $\pm$ 0.1  &       4.16 $\pm$ 0.22  &  	 1.29 $\pm$ 0.49  &  	 3.31 $\pm$ 0.38  &  	 1.83 $\pm$ 0.33  &              10.6 $\pm$ 0.7  $\pm$ 1.2  &         	 15 \\
\vspace{0.13cm}
1150 &    	 13.4 $\pm$ 0.1  & 	 3.05 $\pm$ 0.17  &  	 0.76 $\pm$ 0.38  &	 2.87 $\pm$ 0.35  &  	 1.29 $\pm$ 0.28  &              7.97 $\pm$ 0.61  $\pm$ 0.92  &          13 \\
\vspace{0.13cm}
1200 &    	 9.98 $\pm$ 0.07  &   	 3.02 $\pm$ 0.18  &  	 0.56 $\pm$ 0.32  &	 2.29 $\pm$ 0.31  &  	 1.09 $\pm$ 0.23  &              6.96 $\pm$ 0.54  $\pm$ 0.81  &          11 \\
\vspace{0.13cm}
1250 &    	 7.42 $\pm$ 0.05  &  	 2.68 $\pm$ 0.17  &  	 0.74 $\pm$ 0.37  &  	 2.07 $\pm$ 0.30  &   	 0.59 $\pm$ 0.14  &           	 6.08 $\pm$ 0.52  $\pm$ 0.72  &          11 \\
\vspace{0.13cm}
1300 &    	 5.58 $\pm$ 0.04  &  	 1.61 $\pm$ 0.11  &  	 0.74 $\pm$ 0.37  &  	 1.79 $\pm$ 0.28  &  	 0.73 $\pm$ 0.14  &              4.87 $\pm$ 0.49  $\pm$ 0.55  &          9 \\
\vspace{0.13cm}
1350 &    	 4.21 $\pm$ 0.03  &  	 1.03 $\pm$ 0.07  &	 0.74 $\pm$ 0.37  &  	 1.50 $\pm$ 0.25  &   	 0.70 $\pm$ 0.14  &              3.97 $\pm$ 0.48  $\pm$ 0.43  &          7 \\
\vspace{0.13cm}
1400 &    	 3.19 $\pm$ 0.02  &  	 1.01 $\pm$ 0.08  &	 0.74 $\pm$ 0.37  &  	 1.33 $\pm$ 0.26  &  	 0.69 $\pm$ 0.14  &              3.76 $\pm$ 0.48  $\pm$ 0.39  &          7 \\
\vspace{0.13cm}
1450 &    	 2.42 $\pm$ 0.02  &  	 1.45 $\pm$ 0.12  &  	 0.56 $\pm$ 0.32  &	 1.32 $\pm$ 0.26  &  	 0.65 $\pm$ 0.14  &              3.97 $\pm$ 0.45  $\pm$ 0.44  &          7 \\
\vspace{0.13cm}
1500 &    	 1.84 $\pm$ 0.01  &  	 1.29 $\pm$ 0.11  &  	 0.56 $\pm$ 0.32  &	 1.32 $\pm$ 0.26  &  	 0.58 $\pm$ 0.14  &           	 3.75 $\pm$ 0.45  $\pm$ 0.41  &          7 \\
\vspace{0.13cm}
1550 &    	 1.40 $\pm$ 0.01  &	 1.12 $\pm$ 0.10  &   	 0.56 $\pm$ 0.32  &	 1.32 $\pm$ 0.26  &  	 0.49 $\pm$ 0.14  &           	 3.49 $\pm$ 0.45  $\pm$ 0.39  &          6 \\
\vspace{0.13cm}
1600 &    	 1.07 $\pm$ 0.01  &	 1.07 $\pm$ 0.10  &   	 0.56 $\pm$ 0.32  &	 1.27 $\pm$ 0.26  &  	 0.46 $\pm$ 0.14  &           	 3.35 $\pm$ 0.45  $\pm$ 0.37  &          6 \\
\vspace{0.13cm}
1650 &    	 0.82 $\pm$ 0.01  &	 0.88 $\pm$ 0.09  & 	 0.56 $\pm$ 0.32  &	 1.27 $\pm$ 0.26  &  	 0.44 $\pm$ 0.14  &           	 3.15 $\pm$ 0.44  $\pm$ 0.35  &          6 \\
\vspace{0.13cm}
1700 &    	 0.629 $\pm$ 0.004  &  	 0.99 $\pm$ 0.11  &  	 0.56 $\pm$ 0.32  &	 1.05 $\pm$ 0.24  &  	 0.42 $\pm$ 0.14  &           	 3.01 $\pm$ 0.44  $\pm$ 0.32  &          6 \\
\vspace{0.13cm}
1750 &    	 0.487 $\pm$ 0.003  &  	 0.91 $\pm$ 0.11  &  	 0.38 $\pm$ 0.27  & 	 0.98 $\pm$ 0.23  &  	 0.38 $\pm$ 0.14  &           	 2.65 $\pm$ 0.39  $\pm$ 0.30  &          5 \\
\vspace{0.13cm}
1800 &    	 0.373 $\pm$ 0.002  &  	 0.91 $\pm$ 0.11  &  	 0.38 $\pm$ 0.27  & 	 0.96 $\pm$ 0.24  &  	 0.36 $\pm$ 0.14  &           	 2.61 $\pm$ 0.40  $\pm$ 0.29  &          5 \\
\vspace{0.13cm}
1850 &    	 0.287 $\pm$ 0.002  &  	 0.88 $\pm$ 0.11  &  	 0.20 $\pm$ 0.20  &	 0.90 $\pm$ 0.23  &   	 0.32 $\pm$ 0.14  &           	 2.30 $\pm$ 0.35  $\pm$ 0.28  &          4 \\
\vspace{0.13cm}
1900 &    	 0.221 $\pm$ 0.001  &  	 0.74 $\pm$ 0.10  & 	 0.20 $\pm$ 0.20  &	 0.86 $\pm$ 0.24  &  	 0.31 $\pm$ 0.14  &           	 2.11 $\pm$ 0.35  $\pm$ 0.25  &          3 \\
\vspace{0.13cm}
1950 &    	 0.170 $\pm$ 0.001  &    0.69 $\pm$ 0.10  &	 0.20 $\pm$ 0.20  &	 0.83 $\pm$ 0.24  &  	 0.30 $\pm$ 0.14  &              2.02 $\pm$ 0.35  $\pm$ 0.24  &          3 \\
\vspace{0.13cm}
2000 &    	 0.132 $\pm$ 0.001  &  	 0.68 $\pm$ 0.10  &   	 0.29 $\pm$ 0.20  &	 0.29 $\pm$ 0.09  & 	 0.30 $\pm$ 0.14  &           	 1.47 $\pm$ 0.28  $\pm$ 0.15  &          2 \\
\end{scotch}
}
\label{tab:finalselection_munu}
\end{table*}

\cleardoublepage \section{The CMS Collaboration \label{app:collab}}\begin{sloppypar}\hyphenpenalty=5000\widowpenalty=500\clubpenalty=5000\vskip\cmsinstskip
\textbf{Yerevan Physics Institute, Yerevan, Armenia}\\*[0pt]
A.M.~Sirunyan, A.~Tumasyan
\vskip\cmsinstskip
\textbf{Institut f\"{u}r Hochenergiephysik, Wien, Austria}\\*[0pt]
W.~Adam, F.~Ambrogi, E.~Asilar, T.~Bergauer, J.~Brandstetter, M.~Dragicevic, J.~Er\"{o}, A.~Escalante~Del~Valle, M.~Flechl, R.~Fr\"{u}hwirth\cmsAuthorMark{1}, V.M.~Ghete, J.~Hrubec, M.~Jeitler\cmsAuthorMark{1}, N.~Krammer, I.~Kr\"{a}tschmer, D.~Liko, T.~Madlener, I.~Mikulec, N.~Rad, H.~Rohringer, J.~Schieck\cmsAuthorMark{1}, R.~Sch\"{o}fbeck, M.~Spanring, D.~Spitzbart, A.~Taurok, W.~Waltenberger, J.~Wittmann, C.-E.~Wulz\cmsAuthorMark{1}, M.~Zarucki
% ...
\vskip\cmsinstskip
% ...
1:  Also at Vienna University of Technology, Vienna, Austria\\
% ...
\end{sloppypar}
\end{document}